%% file: main.tex
\documentclass[]{interact}

\usepackage[caption=false]{subfig}
\usepackage[natbibapa,nodoi]{apacite}
\usepackage[utf8]{inputenc}
\usepackage{tabu}
\graphicspath{{img/}}
\usepackage{cases}
\allowdisplaybreaks
\usepackage{siunitx}
\DeclareSIUnit{\decibelm}{dBm}
\DeclareSIUnit{\bits}{bits}
\usepackage[acronym]{glossaries}
\renewcommand{\glossarysection}[2][]{}
\glsdisablehyper
\loadglsentries{text/acronyms} 
\makenoidxglossaries

\usepackage{xcolor}

\begin{document}
\title{Bringing AI to the Edge: A Formal M\&S Specification to Deploy Effective IoT Architectures}
\author{
\name{Román~Cárdenas\textsuperscript{a}, Patricia~Arroba\textsuperscript{a}, and~José L. Risco-Martín\textsuperscript{b}\thanks{CONTACT José L. Risco-Martín. Email: jlrisco@ucm.es}}
\affil{\textsuperscript{a}Universidad Politécnica de Madrid, Madrid, Spain; \
\textsuperscript{b}Universidad Complutense de Madrid, Madrid, Spain}
}

\maketitle

\begin{abstract}
  \input{text/0_abstract}
\end{abstract}

\begin{abbreviations}
\printnoidxglossary[type=\acronymtype]
\end{abbreviations}

\begin{keywords}
Edge Computing; Internet of Things; Computation Offloading; Incremental Learning; Model-Based Systems Engineering
\end{keywords}

\section{Introduction}\label{sec:1_intro}
\input{text/1_introduction}

\section{Related Work}\label{sec:2_related}

\input{text/2_related_work}

\section{System Model and Problem Formulation}\label{sec:3_sysmodel}
\input{text/3_system_model}

\section{Root System Components}\label{sec:4_devmodels}
\input{text/4_dev_models}

\section{Simulation and Results}\label{sec:5_simulation}
\input{text/5_simulation}

\section{Conclusions and Future Work}\label{sec:6_conclusion}
\input{text/6_conclusions}

\section*{Acknowledgments}
\input{text/ack}

\bibliographystyle{apacite}
\bibliography{bibliography}

\appendix
\section{The Discrete Event System Specification}\label{ann:devs}
\input{text/annex_devs}


%
%

\end{document}

%% file: text/acronyms.tex
\newacronym{IoT}{IoT}{Internet of Things}
\newacronym{QoS}{QoS}{Quality of Service}
\newacronym{RTT}{RTT}{Round-Trip Time}
\newacronym{MS}{M\&S}{Modeling and Simulation}
\newacronym{MBSE}{MBSE}{Model-Based Systems Engineering}
\newacronym{DUNIP}{DUNIP}{DEVS Unified Process}
\newacronym{DEVS}{DEVS}{Discrete Event System Specification}
\newacronym{DEVSML}{DEVSML}{DEVS Modeling Language}
\newacronym{ADAS}{ADAS}{Advanced Driver Assistance System}
\newacronym{ISP}{ISP}{Internet Service Provider}
\newacronym{RAN}{RAN}{Radio Access Network}
\newacronym{UE}{UE}{User Equipment}
\newacronym{AP}{AP}{Access Point}
\newacronym{EDC}{EDC}{Edge Data Center}
\newacronym{MDC}{MDC}{Micro Data Center}
\newacronym{NR}{NR}{5G New Radio}
\newacronym{PBCH}{PBCH}{Physical Broadcast Channel}
\newacronym{PSS}{PSS}{Primary Synchronization Signal}
\newacronym{SNR}{SNR}{Signal-to-Noise Ratio}
\newacronym{PRACH}{PRACH}{Physical Random Access Channel}
\newacronym{AMF}{AMF}{Access and Mobility Management Function}
\newacronym{FDD}{FDD}{Frequency Division Duplexing}
\newacronym{PUSCH}{PUSCH}{Physical Uplink Shared Channel}
\newacronym{PDSCH}{PDSCH}{Physical Downlink Shared Channel}
\newacronym{PUCCH}{PUCCH}{Physical Uplink Control Channel}
\newacronym{PDCCH}{PDCCH}{Physical Downlink Control Channel}
\newacronym{MCS}{MCS}{Modulation and Codification Scheme}
\newacronym{RRC}{RRC}{Radio Resource Control}
\newacronym{SysML}{SysML}{Systems Modeling Language}
\newacronym{GPU}{GPU}{Graphics Processing Unit}
\newacronym{CPU}{CPU}{Central Processing Unit}
\newacronym{FSM}{FSM}{Finite State Machine}
\newacronym{DVFS}{DVFS}{Dynamic Voltage and Frequency Scaling}
\newacronym{SDN}{SDN}{Software-Defined Network}
\newacronym{ANN}{ANN}{Artificial Neural Network}
\newacronym{NRMSD}{NRMSD}{Normalized Root Mean Square Deviation}
\newacronym{GPS}{GPS}{Global Positioning System}
\newacronym{EMA}{EMA}{Exponential Moving Average}
\newacronym{QAM}{QAM}{Quadrature Amplitude Modulation}
\newacronym{WLAN}{WLAN}{Wireless Local Area Network}
\newacronym[plural=ABC,firstplural=Abstract Base Classes (ABCs)]{ABC}{ABC}{Abstract Base Class}
\newacronym{DNN}{DNN}{Deep Neural Network}
\newacronym{CNN}{CNN}{Convolutional Neural Network}
\newacronym{CSV}{CSV}{Comma-Separated Values}
\newacronym{M2M}{M2M}{Machine-to-Machine}
\newacronym{LTE}{LTE}{Long Term Evolution}
\newacronym{WAN}{WAN}{Wide Area Network}
\newacronym{V2V}{V2V}{Vehicle-to-Vehicle}
\newacronym{V2I}{V2I}{Vehicle-to-Infrastructure}
\newacronym{MSO}{M\&S\&O}{Modeling, Simulation, and Optimization}
\newacronym{P2P}{P2P}{Point-to-Point}
\newacronym{FaaS}{FaaS}{Function-as-a-Service}
\newacronym{CNF}{CNF}{Core Network Function}
\newacronym{PU}{PU}{Processing Unit}
\newacronym{OSI}{OSI}{Open Systems Interconnection}
\newacronym{FSPL}{FSPL}{Free-Space Path Loss}
\newacronym{FDM}{FDM}{Frequency Division Multiplexing}
\newacronym{ML}{ML}{Machine Learning}
\newacronym{AI}{AI}{Artificial Intelligence}
\newacronym{WSN}{WSN}{Wireless Sensor Network}

%% file: text/0_abstract.tex
The Internet of Things is transforming our society, providing new services that
improve the quality of life and resource management. These applications are based
on ubiquitous networks of multiple distributed devices, with limited computing
resources and power, capable of collecting and storing data from heterogeneous
sources in real-time. To avoid network saturation and high delays, new
architectures such as fog computing are emerging to bring computing
infrastructure closer to data sources. Additionally, new data centers are needed to provide
real-time Big Data and data analytics capabilities at the edge of the network,
where energy efficiency needs to be considered to ensure a sustainable and
effective deployment in areas of human activity.
In this research, we present an IoT model based on the principles of Model-Based
Systems Engineering defined using the Discrete Event System Specification
formalism. The provided mathematical formalism covers the description of the
entire architecture, from IoT devices to the processing units in edge data
centers. Our work includes the location-awareness of user equipment, network, and
computing infrastructures to optimize federated resource management in terms of
delay and power consumption. We present an effective framework to assist the
dimensioning and the dynamic operation of IoT data stream analytics applications,
demonstrating our contributions through a driving assistance use case based on
real traces and data.

%% file: text/1_introduction.tex
Emerging technological advances, such as smart cities,
personalized medicine, and driving assistance systems, among many other
\gls{IoT} applications, are enabling a transformation of our society by
applying \gls{AI} and \gls{ML} algorithms to improve both people's
quality of life and efficient resource management.
This technology is in full growth, and the deployment of 20.41 billion devices
connected to the internet is expected within the next few
years~\citep{GartnerPR2017}.  Also, within the next decade,
McKinsey~\citep{mckinsey2015} estimates that the economic impact of the global
\gls{IoT} market will reach 11.1 trillion USD.

The fog computing paradigm has been designed to support these applications,
which require rapid mobility and intensive data processing in
real-time~\citep{deng2016}. Fog computing, as defined by Cisco
Systems~\citep{cisco2012}, aims to bring the cloud computing paradigm closer to
the edge of the access network, deploying \glspl{EDC} that, together with 5G,
will provide improvements in latency and bandwidth congestion.
\glspl{EDC}, also considered as \glspl{MDC}, are platforms
that provide computation, storage, and networking services to end nodes connected
to an access network~\citep{aazam2015}. In contrast to the cloud, \glspl{EDC} are
extensively distributed, and their resources are dimensioned accordingly to cover
the demand in small areas.
These edge infrastructures may support the massive Big Data requirements
to process and analyze vast amounts of heterogeneous data streams in
real-time to apply
incremental learning techniques (i.e., training current predictive models
with new data to improve the \gls{QoS}~\citep{gepperth2016}).
Solutions based on \gls{GPU} clusters are one of the best
options to provide useful resources for \gls{IoT} applications, as their
architecture allows the efficient training of computationally expensive
\gls{ML} algorithms such as \glspl{ANN}, frequently used in data analytics,
with better performance than CPUs~\citep{8057318}.

Currently, there is no effective deployment and operation of these new
generation \gls{EDC} infrastructures, in terms of energy efficiency, due to
the challenge involved in adapting traditional data centers to the specific
needs in areas of human activity.
The state-of-the-art provides several models and simulators that assist in the
efficient deployment and operation of fog computing infrastructures. However,
the majority of them are software-based frameworks without a background of
mathematical formalism. On the other hand, formal modeling-based solutions are
focused on specific modules and do not cover the entire system, from \gls{IoT}
devices to processing units, which simplifies model management.
Moreover, the complex context of \gls{IoT}, which involves the coordinated
management of user equipment, communication networks, processing units, and
mobility, could benefit from advanced \gls{MS} methods. The solid \gls{MBSE}
principles ensure an incremental design that is logic, robust, and reliable.

In this paper, we propose an \gls{MS} framework for \gls{IoT} scenarios.
The main contributions of this research are the following:
\begin{itemize}
	\item The model is entirely defined using the \gls{DEVS} mathematical
	formalism~\citep{zeigler2000}. \gls{DEVS} provides several advantages, such
	as completeness, verifiability, extensibility, and maintainability.
	\item The approach for modeling the \gls{IoT} applications is based on the
	\gls{FaaS} paradigm. This reactive approach enables better resource optimization
	for scenarios with varying demand. As edge computing is a location-aware
	technology, computation offloading services can improve their performance using
	\gls{FaaS} patterns.
	\item The model contains all the components that comprise the \gls{IoT} scenario.
	Our approach is able to capture interdependencies between elements of the system, providing a more accurate and reliable analysis of the scenario under
	study.
	\item Our work enables the optimization of the \glspl{EDC}' resource
	management in a federated location-aware fashion, in terms of delay and power
	consumption, considering the mobility of users and the placement of both
	network and computing infrastructures. This mathematical formalism helps with
	the energy and delay optimization of edge infrastructures for both the
	deployment dimensioning and the dynamic operation in \gls{IoT} data stream
	applications.
	\item We demonstrate the contributions of our research through a use
	case, using real data for a driving assistance \gls{IoT} application, from
	the location and mobility of devices to processing units' workload and power profiles.
\end{itemize}

%% file: text/2_related_work.tex
In the field of energy efficiency, the majority of \gls{IoT} models based on
mathematical formalisms only cover specific system submodules. Models
formally describing \gls{IoT} devices, communications networks, and access points
can be found in the research of Tom et al.~\citep{8620284}, Zi et
al.~\citep{8733050} and Sheth et al~\citep{8789655}, respectively.

Some models combine several submodules of the complete system.
Verma et al.~\citep{8761998} provide models of IoT devices together with the
network, and Yu et al.~\citep{8675366} design the \gls{IoT} subsystems from the user
equipment to the access points. Research produced by Dong et
al.~\citep{8653370} also includes the computing infrastructure at the edge level,
thus presenting a complete fog computing architecture. However, the model does
not provide user equipment mobility or handovers.

There are software tools for modeling and simulation of specific parts of \gls{IoT}
and edge computing scenarios. For instance, COOJA~\citep{cooja} is a simulator for
\glspl{WSN} that focuses on the firmware of \gls{IoT} devices and the communication
mechanisms between sensors. The FogNetSim++ simulator~\citep{fognetsim}, built on top
of the OMNeT++ network simulator~\citep{omnet}, focuses on connectivity issues of fog
infrastructures, such as end-users mobility, handover processes, and fog communication
protocols. In contrast, the RelIoT simulator~\citep{reliot}, built on top of the ns-3
network simulator~\citep{ns3}, analyzes the reliability of fog nodes of edge computing
scenarios. Alternatively, FogTorch~\citep{fogtorch} is a prototype Monte Carlo-based
simulator that provides an in-depth analysis of the \gls{QoS} of \gls{IoT} applications.

On the other hand, there are complete Fog infrastructure solutions that allow complex
\gls{IoT} applications to be simulated with a wide variety of degrees of freedom.
Among all the available tools, the most popular ones are those developed on top of the
CloudSim simulator~\citep{cloudsim}. First, iFogSim~\citep{ifogsim} adds sensor and
actuator classes to model \gls{IoT} devices. Then, IOTSim~\citep{iotsim} focuses on
optimizing computation offloading services based on the MapReduce
algorithm~\citep{mapreduce}. Finally, EdgeCloudSim~\citep{edgecloudsim} includes simple
nomadic mobility models for end-users. However, all of these developments are
implemented through software simulations that are not supported by mathematical formalisms.

This paper presents the mathematical description and the integration of the
complete \gls{IoT} system, from the user's \gls{IoT} devices to the processing
units of the edge data centers.  Our models follow the principles of \gls{MBSE},
ensuring an incremental design that is logic, robust, and reliable.  Our research
contributes to the state-of-the-art by providing location-awareness, 5G
capabilities, and federated management of the complete infrastructure for
\gls{IoT} applications following the \gls{FaaS} model.

%% file: text/3_system_model.tex
This section first describes a standard edge
computing-based \gls{IoT} architecture. Then, we formulate the
problem and depict a broad vision of the implementation carried
out to represent the model.

\subsection{System Model}\label{subsec:3_1_system}
\begin{figure}[h]
	\centering
	\includegraphics[width=\linewidth]{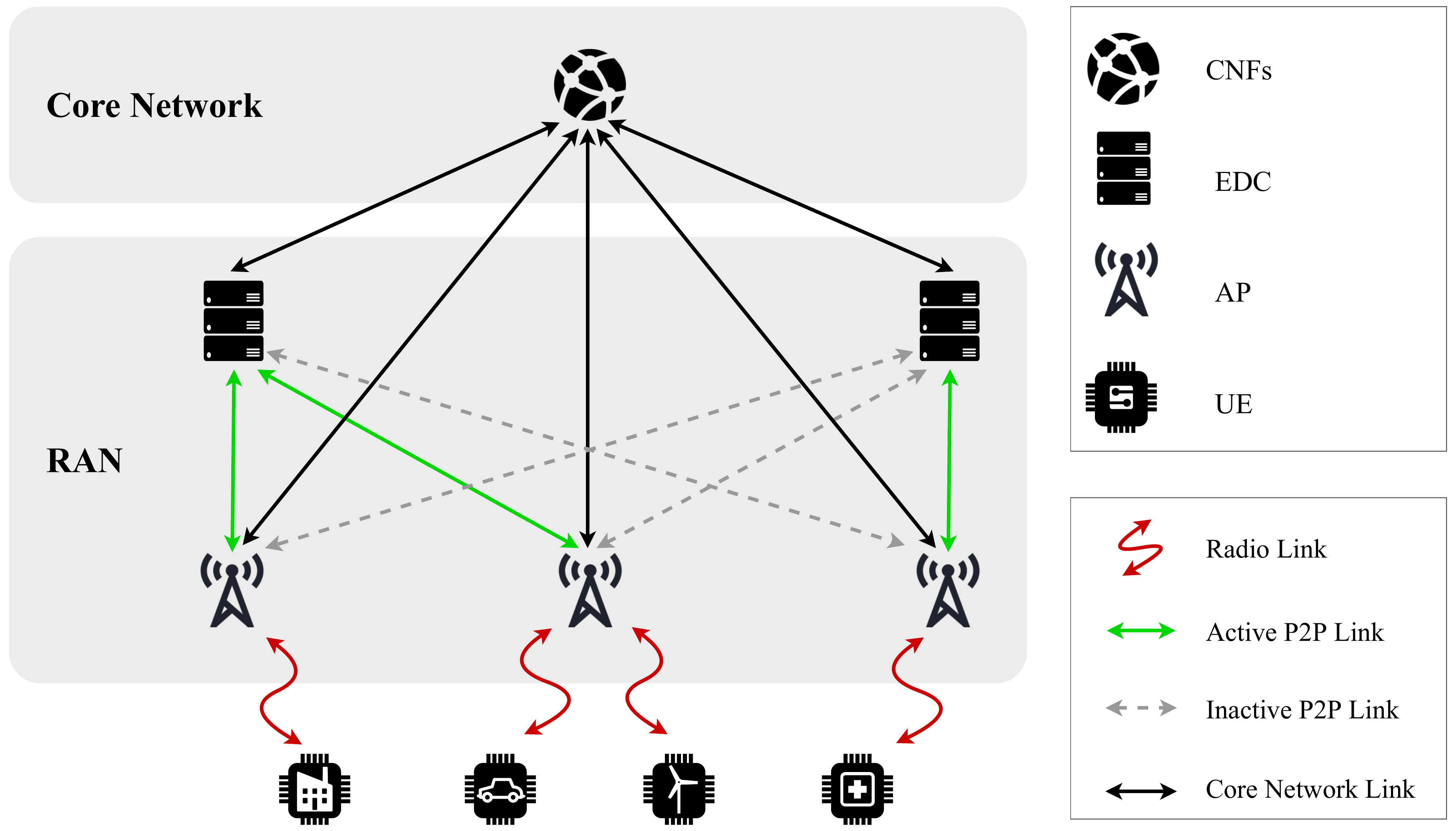}
	\caption{Edge Computing Scenario.}
	\label{fig:scenario}
\end{figure}

We consider an edge computing system with \gls{UE} devices, as
shown in \figurename~\ref{fig:scenario}. 

Each \gls{UE} executes one or more \gls{IoT} services that acquire
data from their environment. Data processing is offloaded to
\glspl{EDC}. \glspl{EDC} are located at the edge of the network,
which is called \gls{RAN}, and have powerful computing capabilities
to improve significantly the quality of computation experience
perceived by each \gls{UE}. \glspl{AP} act as gateways and
establish wireless communication channels between \gls{UE} and
the \gls{RAN}. For enhancing the perceived quality of the signal,
\gls{UE} always connect to their closest \gls{AP}. As \gls{UE} nodes
can change their location, handovers from one \gls{AP} to another
may be triggered. \glspl{AP} forward data packages sent by
\gls{UE} to the \glspl{EDC} via \gls{P2P} optical fiber
communication links.
 
The \gls{ISP} that owns the \gls{RAN} performs different network
management-related tasks using \glspl{CNF}. Concretely, the
\gls{SDN} Control Function monitors the availability of all the
\glspl{EDC} in the \gls{RAN} and activates or deactivates the
\gls{P2P} links between \glspl{EDC} and \glspl{AP} to maximize
\gls{UE}'s perceived \gls{QoS} while ensuring resource management
policies. The \gls{SDN} Control Function introduces dynamic network
slicing features within the \gls{RAN} in a location-aware
manner, activating the \gls{P2P} links between
every \gls{AP} and its closest \gls{EDC} with enough available
computing resources.

The approach followed for modeling \gls{IoT} applications' behavior is
based on the \gls{FaaS} paradigm. No resources are dedicated to any
service in the \gls{EDC}. Only when a \gls{UE} request to open a
new service (i.e., establish a communication link with the \gls{EDC}
to send service-related data) will the \gls{EDC} earmark the required
computing resources for granting computation offloading while the
service is active.

\subsection{Problem Formulation}\label{subsec:3_2_problem}

\begin{figure}[h]
	\centering
	\includegraphics[width=\linewidth]{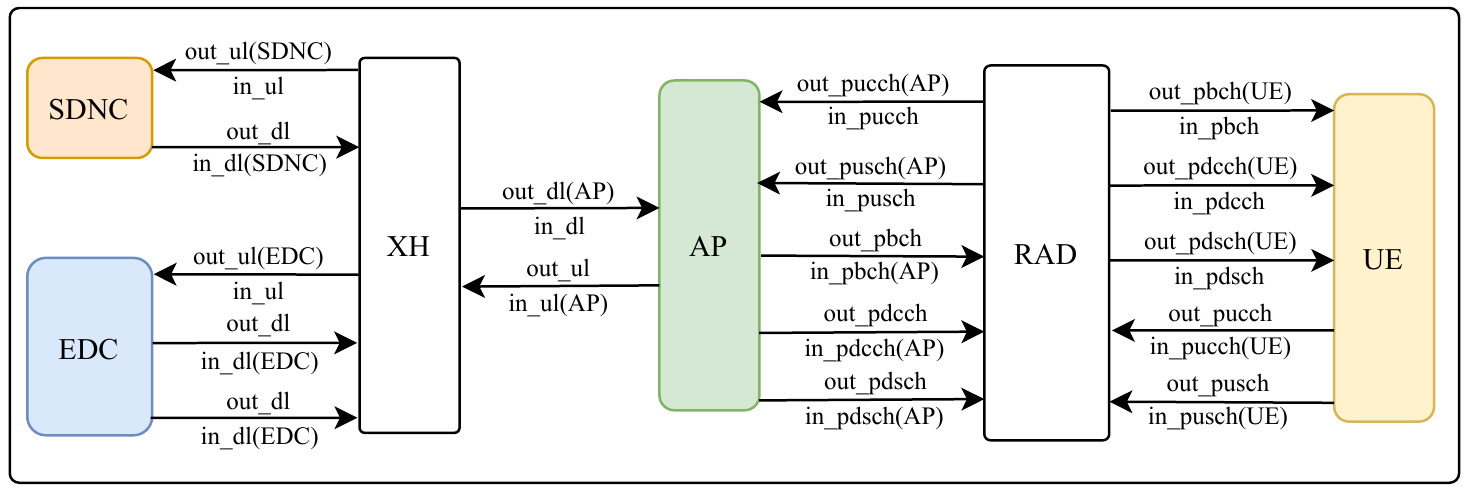}
	\caption{Edge Computing Root System Coupled Model.}
	\label{fig:root}
\end{figure}

For modeling the system under study in a formal way, we have applied
\gls{MBSE} techniques that enabled us to employ analysis,
specification, design, and verification of the system being
developed~\citep{Friedenthal2011}. 
System behavior has been specified following the approach
described by the \gls{DEVS} formalism~\citep{zeigler2000} (see Appendix~\ref{ann:devs}).
In the following, for a given \gls{DEVS} entity $\mathrm{E}$,
we define $\mathbf{E}$ as the set of all the components of this
entity within the scenario. $\mathbf{E}(\mathrm{C})$ is the set
of all the elements of the entity $\mathrm{E}$ that are subcomponents
of the coupled entity $\mathrm{C}$. A simple state variable $\mathrm{s}$
that is part of the state of DEVS model $\mathrm{E}$ is $\mathrm{s(E)}$.
On the other hand, $\mathbf{s}(\mathrm{E})$ is a state variable composed of a set
of variables.
We also denote a connection from port pA of a Component $C_1$ to port pB of a Component
$C_2$ as $C_1 \xrightarrow[\text{pB}]{\text{pA}} C_2$. Hence, the
Coupled \gls{DEVS} model of the root system is:
\begin{align*}
\mathrm{ROOT} &= \langle X,Y,C,EIC,IC,EOC\rangle \\
X &= Y = EIC = EOC = \{\} \\
C &= \{\mathrm{XH},\mathrm{RAD},\mathrm{SDNC},\mathbf{UE},\mathbf{AP},\mathbf{EDC}\} \\
\forall \mathrm{UE} &\in \mathbf{UE},\forall \mathrm{AP} \in \mathbf{AP},\forall \mathrm{EDC} \in \mathbf{EDC}:\\
IC &= \{\mathrm{UE} \xrightarrow[\text{in\_pucch(UE)}]{\text{out\_pucch}} \mathrm{RAD}, \mathrm{UE} \xrightarrow[\text{in\_pusch(UE)}]{\text{out\_pusch}} \mathrm{RAD},\\
	&\hspace{0.6cm}\mathrm{RAD} \xrightarrow[\text{in\_pbch}]{\text{out\_pbch(UE)}} \mathrm{UE}, \mathrm{RAD} \xrightarrow[\text{in\_pdcch}]{\text{out\_pdcch(UE)}} \mathrm{UE},\\
	&\hspace{0.6cm}\mathrm{RAD} \xrightarrow[\text{in\_pdsch}]{\text{out\_pdsch(UE)}} \mathrm{UE}, \mathrm{RAD} \xrightarrow[\text{in\_pucch}]{\text{out\_pucch(AP)}} \mathrm{AP},\\
	&\hspace{0.6cm}\mathrm{RAD} \xrightarrow[\text{in\_pusch}]{\text{out\_pucch(AP)}} \mathrm{AP}, \mathrm{AP} \xrightarrow[\text{in\_pbch(AP)}]{\text{out\_pbch}} \mathrm{RAD},\\
	&\hspace{0.6cm}\mathrm{AP} \xrightarrow[\text{in\_pdcch(AP)}]{\text{out\_pdcch}} \mathrm{RAD}, \mathrm{AP} \xrightarrow[\text{in\_pdsch(AP)}]{\text{out\_pdsch}} \mathrm{RAD},\\
	&\hspace{0.6cm}\mathrm{AP} \xrightarrow[\text{in\_ul(AP)}]{\text{out\_xh}} \mathrm{XH}, \mathrm{XH} \xrightarrow[\text{in\_xh}]{\text{out\_dl(AP)}} \mathrm{AP},\\
	&\hspace{0.6cm}\mathrm{XH} \xrightarrow[\text{in\_ul}]{\text{out\_ul(EDC)}} \mathrm{EDC}, \mathrm{EDC} \xrightarrow[\text{in\_ul(EDC)}]{\text{out\_ul}} \mathrm{XH},\\
	&\hspace{0.6cm}\mathrm{EDC} \xrightarrow[\text{in\_dl(EDC)}]{\text{out\_dl}} \mathrm{XH}, \mathrm{XH} \xrightarrow[\text{in\_ul}]{\text{out\_ul(SDNC)}} \mathrm{SDNC},\\
	&\hspace{0.6cm}\mathrm{SDNC} \xrightarrow[\text{in\_dl(SDNC)}]{\text{out\_dl}} \mathrm{XH}\}
\end{align*}
Note that this Coupled \gls{DEVS} model definition is equivalent
to the schematic shown in \figurename~\ref{fig:root}.
Coupled  \gls{DEVS} models and their corresponding schematics
are used indistinctly for describing entities of the system.


%% file: text/4_dev_models.tex
\subsection{Edge Data Centers ($\mathbf{EDC}$)}
\begin{figure}[h]
	\centering
	\includegraphics[width=\linewidth]{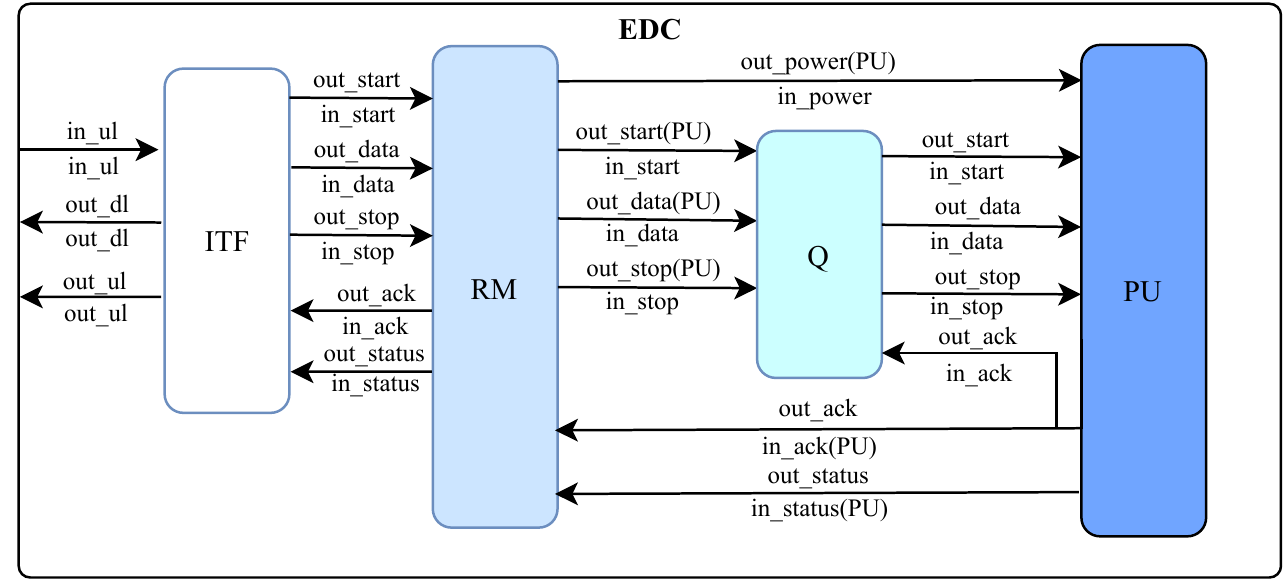}
	\caption{Edge Data Center Coupled Model.}
	\label{fig:edc}
\end{figure}

$\mathrm{EDC}$ models the behavior of the infrastructure deployed for enabling
computation offloading for \gls{UE}. We followed a \gls{FaaS} approach:
\glspl{EDC} are pools of resources. \gls{UE} devices request to start a service
$\mathrm{SRV}$ that requires an \gls{EDC} to process a data stream in real-time.
Depending on the nature of the data to be processed (i.e., the type of \gls{IoT}
application app(SRV)), the service will require a given amount of computing
resources. The \gls{EDC} that handles this request will reserve the requested
amount of computing resources from its pool, and they will no longer be available
for any other service, thus ensuring that the \gls{QoS} expected by the \gls{UE}
will be met once the service starts. When the service is stopped by the \gls{UE},
used resources are available again for any new service request.
The Coupled \gls{DEVS} model for each $\mathrm{EDC}\in\mathbf{EDC}$ is depicted
in \figurename~\ref{fig:edc}.

\subsubsection{Processing Units ($\mathbf{PU}$)}
Each $\mathrm{EDC}$ contains a set of \glspl{PU} $\mathbf{PU}(\mathrm{EDC})$. \glspl{PU} are hardware devices that provide computing capabilities to $\mathrm{EDC}$. Table~\ref{tab:pu} gathers the properties of a \gls{PU}.
\begin{table}[h]
	\caption{Attributes of Processing Units}
	\label{tab:pu}
	\centering
	\begin{tabu} to \linewidth {X[1,c,m] X[4.5,l,m]}
		\bfseries Attribute & \bfseries Definition\\
		\hline\hline
		$t_\mathrm{pw}$ & Time required by the \gls{PU} to power on/off.\\
		$t_\mathrm{srv}$ & Time required by the \gls{PU} to start/stop services.\\
		$t_\mathrm{srv}$ & Time required by the \gls{PU} to start/stop
	services.\\
		$t_\mathrm{data}$ & Time required by the \gls{PU} to process
	incoming messages with data of active services. \\
		$U$ & Overall available computation resources of the \gls{PU}. \\
		\hline
	\end{tabu}
\end{table}
The Atomic \gls{DEVS} model of a $\mathrm{PU}$ component is:
\begin{align*}
		\mathrm{PU}&=\langle X, S, Y,\delta_\mathrm{int},\delta_\mathrm{ext},\lambda,\mathrm{ta}\rangle\\
		X&=\{\mathrm{in\_power},\mathrm{in\_start},\mathrm{in\_data},\mathrm{in\_stop}\}\\
		S&=\{\langle\sigma,\phi,\mathbf{srv},\mathrm{ack}\rangle\}\\
		Y&=\{\mathrm{out\_status},\mathrm{out\_ack}\}\\
		\mathrm{ta}&(\langle\sigma,\phi,\mathbf{srv},\mathrm{ack}\rangle)=\sigma\\
		\lambda &(\langle\sigma,\phi,\mathbf{srv},\mathrm{ack}\rangle)=(\mathrm{out\_status},\mathrm{st}(\mathrm{PU}))\\
		\lambda &(\langle\sigma,\phi,\mathbf{srv},\mathrm{ack}\neq\emptyset\rangle)=(\mathrm{out\_ack},\mathrm{ack})\\
		\delta_\mathrm{int}&(\langle\sigma,\mathrm{to\_on},\emptyset,\emptyset\rangle)=\langle\infty,\mathrm{on},\emptyset,\emptyset\rangle\\
		\delta_\mathrm{int}&(\langle\sigma,\mathrm{to\_off},\emptyset,\emptyset\rangle)=\langle\infty,\mathrm{off},\emptyset,\emptyset\rangle\\
		\delta_\mathrm{int}&(\langle\sigma,\mathrm{busy},\mathbf{srv},\mathrm{ack}\rangle)=\langle\infty,\mathrm{on},\mathbf{srv},\emptyset\rangle\\
		\delta_\mathrm{ext}&(\langle\sigma,\mathrm{off},\emptyset,\emptyset\rangle,e,(\mathrm{in\_power},\mathrm{true}))=\langle t_\mathrm{pw},\mathrm{to\_on},\emptyset,\emptyset\rangle\\
		\delta_\mathrm{ext}&(\langle\sigma,\mathrm{on},\emptyset,\emptyset\rangle,e,(\mathrm{in\_power},\mathrm{false}))=\langle t_\mathrm{pw},\mathrm{to\_off},\emptyset,\emptyset\rangle\\
	\mathrm{Let~}&\mathrm{be~}s=\langle \sigma,\mathrm{on},\mathbf{srv},\emptyset \rangle \\
	\delta_\mathrm{ext}&(s,e,(\mathrm{in\_start},\mathrm{msg}))=\langle t_\mathrm{srv},\mathrm{busy},\mathbf{srv}\cup \mathrm{srv}(\mathrm{msg}),\mathrm{ack}(\mathrm{msg})\rangle\\
	\delta_\mathrm{ext}&(s, e,(\mathrm{in\_stop},\mathrm{msg}))=\langle t_\mathrm{srv},\mathrm{busy},\mathbf{srv}\setminus \mathrm{srv}(\mathrm{msg}),\mathrm{ack}(\mathrm{msg})\rangle\\
	\delta_\mathrm{ext}&(s, e,(\mathrm{in\_data},\mathrm{msg}))=\langle t_\mathrm{data},\mathrm{busy},\mathbf{srv},\mathrm{ack}(\mathrm{msg})\rangle
\end{align*}
$\mathrm{PU}$ can provide service to any $\mathrm{UE}$ only if it is powered.
Once switched on, $PU$ may receive requests for starting a service, injecting new
data to be processed, or stop a service. All the different requests, as well as
the messages to acknowledge them, contain the same fields and are detailed in
Table~\ref{tab:app}.
\begin{table}[h]
	\caption{Fields of IoT Service-Related Messages}
	\label{tab:app}
	\centering
	\begin{tabu} to \linewidth {X[1,c,m] X[4.5,l,m]}
		\bfseries Field & \bfseries Description\\
		\hline\hline
		UE & $\mathrm{UE}\in\mathbf{UE}$ running the service.\\
		APP & $\mathrm{APP}\in\mathbf{APP}$, Application type of the service.\\
		EDC & $\mathrm{EDC}\in\mathbf{EDC}$ that performs computation offloading.\\
		U & Amount of reserved computing resources.\\
		S & Message size. All the messages have 0 bits, except from data messages, which size depends on the application.\\
		\hline
	\end{tabu}
\end{table}
Each $\mathrm{PU}$ hosts a set of active services $\mathbf{srv}$.
The utilization of computing resources of $\mathrm{PU}$ at time t can
never be higher than its total available resources:
\begin{equation}
	u(\mathrm{PU})=\sum U(\mathrm{srv}\in\mathbf{srv}(\mathrm{PU}))\leq U(\mathrm{PU})
\end{equation}
Processing units consume power for providing computation
offloading. Power consumption $\mathrm{pw}(\mathrm{PU})$ depends on the
immediate utilization of its hardware resources.
$\mathrm{PU}$ send periodic reports of their status via the $\mathrm{out\_status}$ port:
\begin{equation}
	\mathrm{st}(\mathrm{PU})=\langle\mathbf{srv}(\mathrm{PU}),\mathrm{pw}(\mathrm{PU}),u(\mathrm{PU}),U(\mathrm{PU})\rangle
\end{equation}

\subsubsection{Service Queues ($\mathbf{Q}$)}
There is a service queue for each \gls{PU}. The queue forwards
service-related messages one by one to the \gls{PU}. While the \gls{PU}
is processing any request, the queue stores any incoming request in a
buffer. When the \gls{PU} sends a response to a processed request, the
service queue forwards the next message of the queue to the \gls{PU}.

The atomic \gls{DEVS} model of a service queue is:

\begin{align*}
		\mathrm{Q}&=\langle X, S, Y,\delta_\mathrm{int},\delta_\mathrm{ext},\lambda,\mathrm{ta}\rangle\\
		X&=\{\mathrm{in\_start},\mathrm{in\_data},\mathrm{in\_stop}, \mathrm{in\_ack}\}\\
		S&=\{\langle\mathrm{busy},\mathbf{stop},\mathbf{start},\mathbf{data}\rangle\}\\
		Y&=\{\mathrm{out\_start},\mathrm{out\_data},\mathrm{out\_stop}\}\\
		\mathrm{ta}&(\langle 1,\mathbf{stop},\mathbf{start},\mathbf{data}\rangle)=\infty\\
		\mathrm{ta}&(\langle 0,\mathbf{stop},\mathbf{start},\mathbf{data}\rangle)=\begin{cases}
			\infty,&\text{if }\mathbf{start}=\mathbf{data}=\mathbf{stop}=\emptyset\\
			0,&\text{otherwise}\\
		\end{cases}\\\\
		\lambda &(\langle 0,\mathbf{stop}\neq\emptyset,\mathbf{start},\mathbf{data}\rangle)=(\mathrm{out\_stop},\mathbf{stop}\{0\}))\\
		\lambda &(\langle 0,\emptyset,\mathbf{start}\neq\emptyset,\mathbf{data}\rangle)=(\mathrm{out\_start},\mathbf{start}\{0\}))\\
		\lambda &(\langle 0,\emptyset,\emptyset,\mathbf{data}\neq\emptyset\rangle)=(\mathrm{out\_data},\mathbf{data}\{0\}))\\
		\delta_\mathrm{int}&(\langle 0,\mathbf{stop}\neq\emptyset,\mathbf{start},\mathbf{data}\rangle)=\langle 1,\mathbf{stop}\setminus\mathbf{stop}\{0\},\mathbf{start},\mathbf{data}\rangle\\
		\delta_\mathrm{int}&(\langle 0,\emptyset,\mathbf{start}\neq\emptyset,\mathbf{data}\rangle)=\langle 1,\emptyset,\mathbf{start}\setminus\mathbf{start}\{0\},\mathbf{data}\rangle\\
		\delta_\mathrm{int}&(\langle 0,\emptyset,\emptyset,\mathbf{data}\neq\emptyset\rangle)=\langle 1,\emptyset,\emptyset,\mathbf{data}\setminus\mathbf{data}\{0\}\rangle\\
		\mathrm{Let~}&\mathrm{be~}s=\langle\mathrm{busy},\mathbf{stop},\mathbf{start},\mathbf{data}\rangle \\
		\delta_\mathrm{ext}&(s,e,(\mathrm{in\_ack},\mathrm{res}))=\langle 0,\mathbf{stop},\mathbf{start},\mathbf{data}\rangle\\
		\delta_\mathrm{ext}&(s,e,(\mathrm{in\_stop},\mathrm{msg}))=\langle \mathrm{busy},\mathbf{stop}\cup\mathrm{msg},\mathbf{start},\mathbf{data}\rangle\\
		\delta_\mathrm{ext}&(s,e,(\mathrm{in\_start},\mathrm{msg}))=\langle \mathrm{busy},\mathbf{stop},\mathbf{start}\cup\mathrm{msg},\mathbf{data}\rangle\\
		\delta_\mathrm{ext}&(s,e,(\mathrm{in\_data},\mathrm{msg}))=\langle \mathrm{busy},\mathbf{stop},\mathbf{start},\mathbf{data}\cup\mathrm{msg}\rangle\\
\end{align*}

\subsubsection{Resource Manager ($RM$)}
This module administrates the \glspl{PU} and decides which \gls{PU}
is to perform the computation offloading of new incoming services
according to a given dispatching algorithm, $PU_{\text{next}}(SRV)$.

$\mathrm{RM}$ also controls which \glspl{PU} must be powered on.
Every \gls{PU} with ongoing services must be turned on. $\mathrm{RM}$
will also keep powered $N_{STBY}$ processing units with no active
services. Keeping \glspl{PU} in hot standby reduces the average
perceived delay while increasing the power consumption of the \gls{EDC}.

Also, $\mathrm{RM}$ aggregates the status reports of all
$\mathrm{PU}\in\mathbf{PU}(\mathrm{EDC})$ to comprise the operational
status of the $\mathrm{EDC}$:
\begin{equation}\label{eqn:st_edc1}
	\begin{aligned}
		\mathbf{srv}(\mathrm{EDC})=\bigcup\mathbf{srv}(\mathrm{PU}\in\mathbf{PU}(\mathrm{EDC}))\\
		\mathrm{pw}(\mathrm{EDC})=\sum \mathrm{pw}(\mathrm{PU}\in\mathbf{PU}(\mathrm{EDC}))\\
		u(\mathrm{EDC})=\sum u(\mathrm{PU}\in\mathbf{PU}(\mathrm{EDC}))\\
		U(\mathrm{EDC})=\sum U(\mathrm{PU}\in\mathbf{PU}(\mathrm{EDC}))\\
	\end{aligned}
\end{equation}

Parameters described in
Eq.~(\ref{eqn:st_edc1}) compound
the state of the \gls{EDC}, $\mathrm{st}(\mathrm{EDC})$,
and are sent to the $\mathrm{SDNC}$ whenever a change is produced.

\subsubsection{Interface ($\mathrm{ITF}$)}
The interface interacts with other components of the root system.
It decapsulates incoming physical messages and forwards their
content to $\mathrm{RM}$. Inversely, messages from $\mathrm{RM}$ are encapsulated
as physical messages and sent to their receiver through $\mathrm{XH}$.

\subsection{Software-Defined Network Control Function ($\mathrm{SDNC}$)}

The \gls{SDN} Control Function monitors the availability of the
\glspl{EDC} within the \gls{RAN} and activates or deactivates
the \gls{P2P} links between \glspl{EDC} and \glspl{AP} to minimize
\gls{UE}' perceived delay while ensuring resource management
policies required by the \gls{ISP}. 
Its atomic \gls{DEVS} definition is:
\begin{align*}
	\mathrm{SDNC}&=\langle X,S,Y,\delta_\mathrm{int},\delta_\mathrm{ext},\lambda,\mathrm{ta}\rangle\\
	X&=\{\mathrm{in\_ul}\}\\
	S&=\{\langle\sigma,\mathbf{st}\rangle\}\\
	Y&=\{\mathrm{out\_dl}\}\\
	\mathrm{ta}&(\langle\sigma,\mathbf{st}\rangle)=\sigma\\
	\lambda&(\langle\sigma,\mathbf{st}\rangle)=(\mathrm{out\_dl},\mathrm{edc(AP)}),\forall \mathrm{AP}\in\mathbf{AP}\\
	\delta_\mathrm{int}&(\langle\sigma,\mathbf{st}\rangle)=\langle\infty,\mathbf{st}\rangle\\
	\delta_\mathrm{ext}&(\langle\sigma,\mathbf{st}\rangle,e,(\mathrm{in\_ul},\mathrm{st}(\mathrm{EDC})))=\langle0,\mathbf{st'}\rangle
\end{align*}


$\mathrm{SDNC}$ keeps a record with the latest status of every
$\mathrm{EDC}$ in $\mathbf{st}$.
Whenever a new \gls{EDC} status is received via the $\mathrm{in\_ul}$ port,
$\mathrm{SDNC}$ assigns to each  $\mathrm{AP}\in\mathbf{AP}$ the closest 
$\mathrm{EDC}$ with enough available  resources:
\begin{equation}
	\begin{aligned}
		\forall\thinspace \mathrm{AP}\in\mathbf{AP}:&\\
		\mathrm{edc}(\mathrm{AP})=\underset{\mathrm{EDC}\in\mathbf{EDC}}{\text{arg min}}&d(\mathrm{AP},\mathrm{EDC})\\
		\text{st. }\hspace{0.3cm}&u(\mathrm{EDC}) < U(\mathrm{EDC})\\
	\end{aligned}
\end{equation}
Assignations are sent to each $\mathrm{AP}\in\mathbf{AP}$ via the  $\mathrm{out\_dl}$ port.

\subsection{Access Points ($\mathbf{AP}$)}\label{subsec:ap}

\begin{figure}[h]
	\centering
	\includegraphics[width=\linewidth]{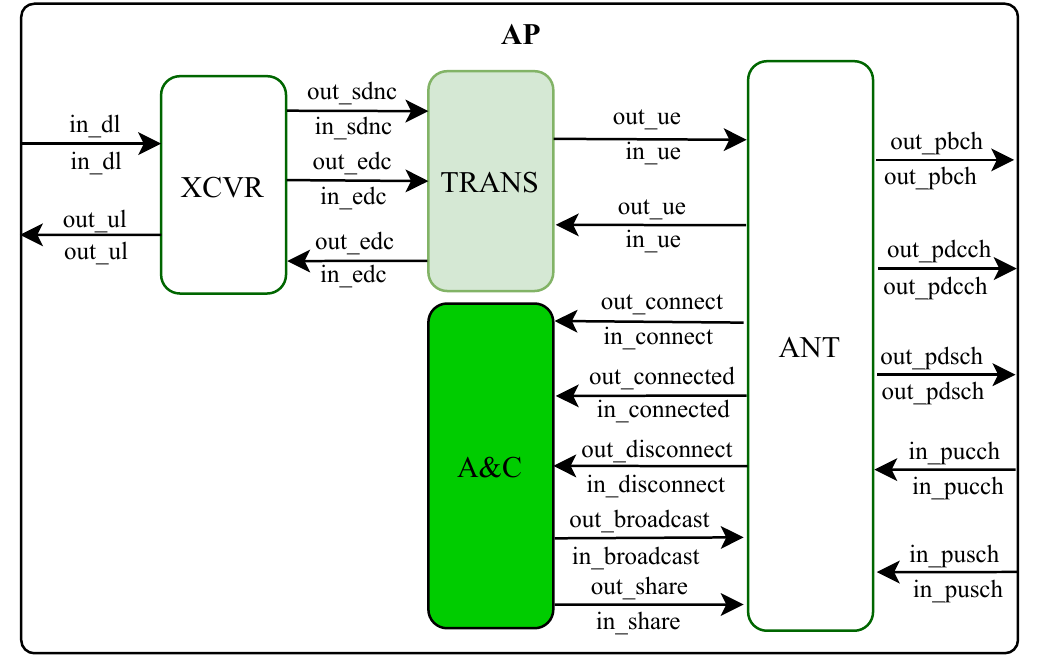}
	\caption{Access Point Coupled Model.}
	\label{fig:ap}
\end{figure}

\glspl{AP} act as gateways between \gls{UE} devices and the rest of
the scenario. For each $AP\in\mathbf{AP}$, its coupled \gls{DEVS}
model is depicted in \figurename~\ref{fig:ap}.

\subsubsection{Access and Control ($A\&C$)}
The Access and Control module is devoted to managing \gls{UE}
connectivity issues. Its Atomic \gls{DEVS} model is:
\begin{align*}
		A\&C&=\langle X,S,Y,\delta _{int},\delta _{ext},\lambda,ta\rangle\\
		X&=\{in\_connect,in\_connected,in\_disconnect\}\\
		S&=\{\langle\sigma,\mathbf{share}\rangle\}\\
		Y&=\{out\_broadcast,out\_share\}\\
		ta&(\langle\sigma,\mathbf{share}\rangle)=\sigma\\
		\lambda&(\langle\sigma,\mathbf{share}\rangle)=\{(out\_share,share(UE)\in\mathbf{share}),(out\_broadcast,pss(AP)\}\\
		\delta _{int}&(\langle\sigma,\mathbf{share}\rangle)=\langle t_{pss},\mathbf{share}\rangle\\
		\delta _{ext}&(\langle\sigma,\mathbf{share}\rangle,e,(in\_port\in X,(UE,\text{report}))=\langle0,\mathbf{share'}\rangle\\
\end{align*}
This module sends \gls{PSS} messages every $t_{pss}$ seconds via
the $out\_broadcast$ port, enabling every $UE\in\mathbf{UE}$
to estimate which \gls{AP} offers to it the best signal quality.
$A\&C$ keeps a record of which $UE\in\mathbf{UE}$ is connected
to the \gls{AP}. New connections are reported via the $in\_connect$
port. Inversely, a new message is received through the $in\_disconnect$
port when a \gls{UE} disconnects.

On the other hand, $A\&C$ is responsible for setting the spectral
efficiency and bandwidth share of connected \gls{UE} nodes:
communication between \glspl{AP} and \gls{UE} share physical
means, and therefore, the available radio spectrum for
communication is divided in a \gls{FDD} manner. Furthermore,
the perceived \gls{SNR} is different for each \gls{UE}, and the
spectral efficiency may vary.

$A\&C$ receives reports regarding the connection quality of connected
\gls{UE} via the $in\_connected$ port. Incoming messages
$report(UE)=\langle SNR_{DL},SNR_{UL}\rangle$
give information about the downlink \gls{SNR} perceived by $UE$ 
and uplink \gls{SNR} perceived by the $AP$ for messages from $UE$.

For each direction of the communication, $(A\&C)$ computes the
maximum theoretical capacity per bandwidth according to
Shannon-Hartley theorem~\citep{shannon1948}:
\begin{equation}
	\frac{C}{{B}_{XL}}(UE)=\log _2(1+SNR_{XL}(UE))
\end{equation}
Depending on the theoretical limit, spectral efficiency is
selected for both uplink and downlink. \glspl{MCS} tables
$\mathbf{MCS}_{UL}$ and $\mathbf{MCS}_{DL}$ contain all the
possible \glspl{MCS} (and their corresponding spectral
efficiency) for each direction of the link. As our radio model
is based on 5G, we used the \gls{MCS} tables defined by the
3GPP in the \gls{NR} standard~\citep{3gpp38214}. This table
contains 28 different \gls{MCS} with spectral efficiencies
ranging from $0.2344$ to $7.4063$ \si{\bits\per\second\per\hertz}
for uplink, and 29 for downlink with spectral efficiencies ranging
from $0.2344$ to $5.5547$ \si{\bits\per\second\per\hertz}.

Finally, the maximum spectral efficiency that fulfills 
the Shannon-Hartley theorem is chosen from $\mathbf{MCS}_{UL}$ and
$\mathbf{MCS}_{DL}$:
\begin{equation}
	\begin{aligned}
		\text{eff}_{XL}(UE)=\text{max}&\{MCS\in\mathbf{MCS}_{XL}\}\\
		\text{st. }&MCS \leq\frac{C}{B}_{XL}(UE)\\
	\end{aligned}
\end{equation}
The set $\mathbf{ue}(AP)\subseteq\mathbf{UE}$ represents all the \gls{UE} devices of the scenario that are connected to the network via $AP$. Every $UE\in\mathbf{ue}(AP)$ gets assigned a bandwidth share.
Radio bandwidth assignation is inversely proportional to $eff_{UL}(UE)$:
\begin{equation}
	share(UE)=\left(\text{eff}_{UL}(UE)\times\sum_{ue\in\mathbf{ue}(AP)}\frac{1}{\text{eff}_{UL}(ue)}\right)^{-1}
\end{equation}
This strategy assigns more bandwidth to those \gls{UE} nodes with lower
spectral efficiency to compensate for this difference and provide a uniform
\gls{QoS} for every connected \gls{UE}.

For every connected \gls{UE}, 
$\mathbf{share}(UE)=\langle bw\_share,\text{eff}_{UL},\text{eff}_{DL}\rangle$
contains information of the bandwidth share and espectral efficiency
of both uplink and downlink communications.
Updated values of $share(UE)\in\mathbf{share}$ are sent to the
$ANT$ component of $AP$ through the $out\_ue$ port.

\subsubsection{Transport ($TRANS$)}
This component forwards messages between \glspl{EDC} and 
\gls{UE} connected to the \gls{AP}. It is modeled as
the following Atomic \gls{DEVS} module:
\begin{align*}
		TRANS&=\langle X,S,Y,\delta _{int},\delta _{ext},\lambda,ta\rangle\\
		X&=\{in\_sdnc,in\_edc,in\_ue\}\\
		S&=\{\langle\sigma,\mathbf{edc},\mathbf{to\_edc},\mathbf{to\_ue}\rangle \}\\
		Y&=\{out\_edc,out\_ue\}\\
		ta&(\langle\sigma,\mathbf{edc},\mathbf{to\_edc},\mathbf{to\_ue}\rangle)=\sigma\\
		\lambda &(\langle\sigma,\mathbf{edc},\mathbf{to\_edc},\mathbf{to\_ue}\rangle)=\{(out\_edc,\mathbf{to\_edc}),(out\_ue,\mathbf{to\_ue})\}\\
		\delta _{int}&(\langle\sigma,\mathbf{edc},\mathbf{to\_edc},\mathbf{to\_ue}\rangle)=\langle\infty,\mathbf{edc},\emptyset,\emptyset\rangle\\
		\delta _{ext}&(\langle\sigma,\mathbf{edc},\mathbf{to\_edc},\mathbf{to\_ue}\rangle,e,(in\_sdnc,\mathbf{edc'}))=\langle0,\mathbf{edc'},\mathbf{to\_edc},\mathbf{to\_ue}\rangle\\
		\delta _{ext}&(\langle\sigma,\mathbf{edc},\mathbf{to\_edc},\mathbf{to\_ue}\rangle,e,(in\_edc,msg))=\langle0,\mathbf{edc},\mathbf{to\_edc},\mathbf{to\_ue}\cup msg\rangle\\
		\delta _{ext}&(\langle\sigma,\mathbf{edc},\mathbf{to\_edc},\mathbf{to\_ue}\rangle,e,(in\_ue,msg))=\langle0,\mathbf{edc},\mathbf{to\_edc}\cup msg,\mathbf{to\_ue}\rangle\\
\end{align*}
The state variable $\mathbf{edc}$ is the offloading routing table. 
$\mathbf{edc}\in\mathbf{EDC}$
corresponds to the \gls{EDC} designated by the $SDNC$ to start
new service from $AP$. 
In case a $UE$ sends a message to start new service $SRV$,
the request
is forwarded to the $\mathbf{edc}$. On the other hand,
requests for stopping a service or messages with new data
for an ongoing service are forwarded to the
$EDC\in\mathbf{EDC}$ that is performing the computation
offloading, regardless of $\mathbf{edc}$.

Variables $\mathbf{to\_edc}$ and $\mathbf{to\_ue}$ are buffers used
by $TRANS$ to temporally store messages to be forwarded to an
$EDC$ or a $UE$, respectively.

\subsubsection{Transceiver ($XCVR$)}\label{subsubsec:xcvr}
$XCVR$ models an optical fiber transceiver used for communicating
with elements of the \gls{RAN}. All the messages $msg$ from $TRANS$
that are to  be sent to  $EDC\in\mathbf{EDC}$ via $XH$ are
encapsulated as physical  messages $M$ and sent through the $out\_xh$
port. Inversely, physical messages $M$ received from $in\_xh$ are
decapsulated and sent to the corresponding port of $TRANS$.

\subsubsection{Antenna ($ANT$)}\label{subsubsec:ap_ant}
$ANT$ is the antenna used by the \gls{AP} for communicating
with \gls{UE} via wireless channels. The attributes of
an antenna are described in Table~\ref{tab:antenna}.
\begin{table}[h]
	\caption{Attributes of a Radio Antenna}
	\label{tab:antenna}
	\centering
	\begin{tabu} to \linewidth {X[1,c,m] X[4.5,l,m]}
		\bfseries Attribute & \bfseries Definition\\
		\hline\hline
		$BW$ & Total available bandwidth per radio channel. It is set to 100 \si{\mega\hertz}.\\
		$PW(ANT)$ & Antenna's transmitting power. It is set to 50 \si{\decibelm}.\\
		$G(ANT)$ & Antenna's gain. It is set to 0 \si{\decibel}.\\
		$T_e(ANT)$ & Antenna's equivalent noise temperature, set to 300 \si{\kelvin}.\\
		\hline
	\end{tabu}
\end{table}

$ANT$ keeps the latest $\mathbf{share}$ sent by the $A\&C$.
The share is sent to the corresponding $UE$ via the $out\_pdcch$
port with to notify the connected \gls{UE} to use the
latest assigned share and spectral efficiencies.

All the messages $msg$ from either $A\&C$ or $TRANS$ that are to
be sent to $UE\in\mathbf{UE}$ via the $out\_pdcch$ or $out\_pdsch$
ports are encapsulated as physical messages $M$ and sent to the
corresponding radio channel output port, setting the bandwidth
$bw(M)$ and spectral efficiency $eff(M)$ to the latest
$\mathbf{share}$, the transmitted power $pw(M)$ to  $P+G$,
and the message size $s(M)$ to the size of the data to be
transmitted $s(msg)$.

\gls{PSS} messages generated by $A\&C$ are sent to all the
$UE\in\mathbf{UE}$ through the $out\_pbch$ using $BW$ \si{\hertz} and
1 \si{\bits\per\second\per\hertz}.
Inversely, physical messages received from $in\_pucch$ input port are
forwarded to either the $in\_connect$ or $in\_disconnect$ ports of $A\&C$.
Messages received from $in\_pusch$ are sent to the $TRANS$ component.
Furthermore, for every physical message $M$ received from a connected
$UE\in\mathbf{share}$, the \gls{SNR} is computed and reported to the
$A\&C$ component via its $in\_connected$ port:
\begin{equation}
	SNR=\frac{P_{W}(M)}{k\times T_e\times bw(M)}
\end{equation}
Where $P_{W}(M)$ is $pw(M)$ converted from \si{\decibelm} to watts:
\begin{equation}
	P_{W}(M)=10^{\frac{pw(M)-30}{10}}
\end{equation}

\subsection{User Equipments ($UE$)}
\begin{figure}[h]
	\centering
	\includegraphics[width=0.9\linewidth]{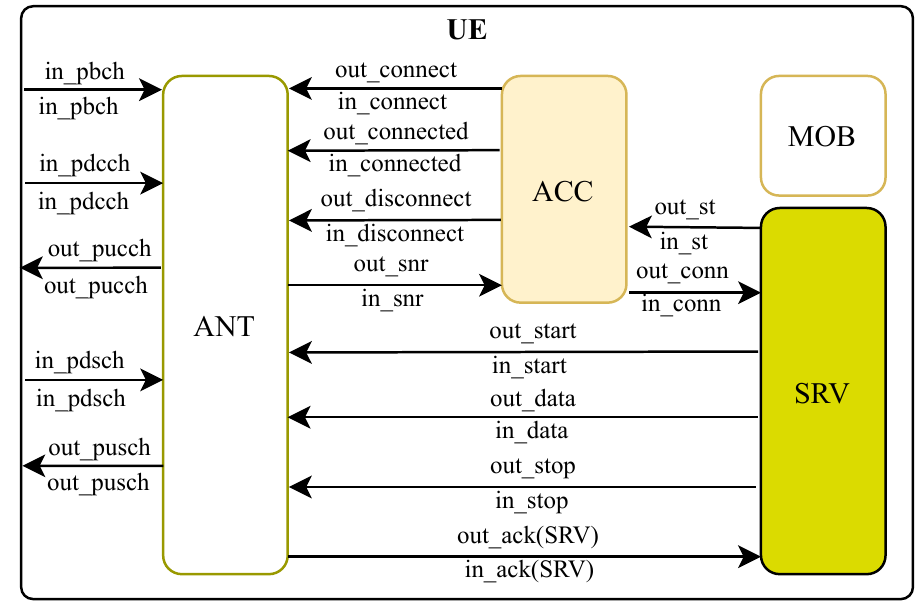}
	\caption{User Equipment Coupled Model.}
	\label{fig:ue}
\end{figure}

The coupled model of each $UE\in\mathbf{UE}$ is depicted in
\figurename~\ref{fig:ue}.

\subsubsection{Service ($SRV$)}
Each \gls{UE} executes a set of \gls{IoT} services
$\mathbf{SRV}(UE)\subseteq\mathbf{SRV}$. These services require
computation offloading for improving its performance.
The attributes of a service are described in
Table~\ref{tab:srv}.
\begin{table}[h]
	\caption{Attributes of IoT Services}
	\label{tab:srv}
	\centering
	\begin{tabu} to \linewidth {X[1,c,m] X[4.5,l,m]}
		\bfseries Attribute & \bfseries Definition\\
		\hline\hline
		$app(SRV)$ & Type of \gls{IoT} application run by $SRV$.\\
		$u(SRV)$ & Required computation resources for providing computation offloading to $SRV$.\\
		$t_{off}(SRV)$ & Inactive time (in seconds). \\
		$t_{on}(SRV)$ & Active time (in seconds). \\
		$s(SRV)$ & Size (in bits) of data messages of $SRV$.\\
		$t_{pkg}(SRV)$ & Message packaging period: every $t_{pkg}(SRV)$ seconds, $SRV$ generates a new data message.\\
		\hline
	\end{tabu}
\end{table}

Each service $SRV$ is described as the following Coupled \gls{DEVS} model:
\begin{align*}
		SRV&=\langle X,Y,C,EIC,IC,EOC\rangle\\
		X&=\{in\_conn,in\_ack\}\\
		Y&=\{out\_st,out\_start,out\_stop,out\_data\}\\
		C&=\{GEN,MNG\}\\
		EIC&=\{SRV \xrightarrow[\text{in\_conn}]{\text{in\_conn}} MNG, SRV \xrightarrow[\text{in\_ack}]{\text{in\_ack}} MNG\}\\
		IC&=\{GEN \xrightarrow[\text{in\_data}]{\text{out}} MNG\}\\
		EOC&=\{MNG \xrightarrow[\text{out\_st}]{\text{out\_st}} SRV, MNG \xrightarrow[\text{out\_start}]{\text{out\_start}} SRV,\\
			&\hspace{0.6cm}MNG \xrightarrow[\text{out\_stop}]{\text{out\_stop}} SRV,MNG \xrightarrow[\text{out\_data}]{\text{out\_data}} SRV\}
\end{align*}

Every $t_{pkg}(SRV)$ seconds, the data generator component $GEN$
generates a new data package of size $s(SRV)$:
\begin{align*}
		GEN&=\langle X,S,Y,\delta _{int},\delta _{ext},\lambda,ta\rangle\\
		X&=\{\}\\
		S&=\{\sigma\}\\
		Y&=\{out\}\\
		ta(\sigma)&=\sigma\\
		\lambda(\sigma)&=(out,s(SRV))\\
		\delta _{int}(\sigma)&=t_{pkg}(SRV)\\
\end{align*}

The service manager $MNG$ is in charge of managing computation
offloading for $SRV$. Its behavior is described as the following
Atomic \gls{DEVS} model:
\begin{align*}
		MNG&=\langle X,S,Y,\delta _{int},\delta _{ext},\lambda,ta\rangle\\
		X&=\{in\_data,in\_conn,in\_ack\}\\
		S&=\{\langle\sigma,\phi,edc\in\mathbf{EDC},\mathbf{q},t\rangle\}\\
		Y&=\{out\_st,out\_start,out\_stop,out\_data\}\\
		ta&(\langle\sigma,\phi,edc,\mathbf{q},t\rangle)=\sigma\\
		\lambda &(\langle\sigma,off,edc,\mathbf{q},t\rangle)=(out\_st,\langle app,true\rangle)\\
		\lambda &(\langle\sigma,wait\_conn,\emptyset,\mathbf{q},t\rangle)=(out\_start,\langle UE,app,u\rangle)\\
		\lambda &(\langle\sigma,send\_data,edc,\mathbf{q},t\rangle)=\begin{cases}
			(out\_stop,\langle UE,app,EDC\rangle),&\text{if }t\leq\sigma\\
			(out\_data,\langle UE,app,edc,\mathbf{q}\{0\}\rangle),&\text{otherwise}
		\end{cases}\\
		\lambda &(\langle\sigma,wait\_stop,edc,\mathbf{q},t\rangle)=(out\_st,\langle app,true\rangle)\\
		\delta _{int}&(\langle\sigma,off,\emptyset,\mathbf{q},t\rangle)=\langle\infty,wait\_conn,\emptyset,\mathbf{q},t_{on}\rangle\\
		\delta _{int}&(\langle\sigma,wait\_conn,\emptyset,\mathbf{q},t\rangle)=\langle\infty,wait\_start,\emptyset,\mathbf{q},t\rangle\\
		\delta _{int}&(\langle\sigma,send\_data,edc,\mathbf{q},t\rangle)=\begin{cases}
			\langle\infty,wait\_stop,edc,\mathbf{q},0\rangle,&\text{if } t\leq\sigma\\
			\langle\infty,wait\_ack,edc,\mathbf{q},t-\sigma\rangle,&\text{if } t>\sigma,\mathbf{q}\neq\emptyset\\
			\langle t-\sigma,send\_data,edc,\mathbf{q},t-\sigma\rangle,&\text{otherwise}\\
		\end{cases}\\
		\delta _{int}&(\langle\sigma,wait\_stop,edc,\mathbf{q},t\rangle)=\langle t_{off},off,\emptyset,\mathbf{q},t_{off}\rangle\\
		\delta _{ext}&(\langle\sigma,\phi,edc,\mathbf{q},t\rangle,e,(in\_data,s))=\langle\sigma-e,\phi,edc,\mathbf{q}\leftarrow s,t-e\rangle\\
		\delta _{ext}&(\langle \sigma,send\_data,\mathbf{q},t\rangle,e,(in\_data,s))=\langle 0,send\_data,\mathbf{q}\leftarrow msg,t-e\rangle\\
		\delta _{ext}&(\langle \sigma,wait\_conn,\emptyset,\mathbf{q},t\rangle,e,(in\_conn,true))=\langle 0,wait\_conn,\emptyset,\mathbf{q},t-e\rangle\\
		\delta _{ext}&(\langle \sigma,wait\_start,\emptyset,\mathbf{q},t\rangle,e,(in\_ack,EDC))=\langle 0,send\_data,EDC,\mathbf{q},t-e\rangle\\
		\delta _{ext}&\langle \sigma,wait\_ack,edc,\mathbf{q},t\rangle,e,(in\_ack,edc))=\langle 0,send\_data,edc,\mathbf{q}\setminus\mathbf{q}\{0\},t-e\rangle\\
\end{align*}
New messages generated by $GEN$ are stored in a queue $\mathbf{q}$.
After $t_{off}(SRV)$ seconds, $MNG$ notifies that it is active,
and waits for the \gls{UE} to be connected to the \gls{RAN}. Once
connected, $SRV$ sends a request for starting a service (i.e.,
computation offloading is requested). When the service is
successfully started, $SRV$ sends the oldest message stored in
$\mathbf{q}$. Once a sent message is acknowledged via
a $\delta _{ext}$ transition triggered by port $in\_msg\_ack$,
acknowledged message $msg$ is removed from $\mathbf{q}$,
and the next message is sent. After $t_{on}(SRV)$ seconds being
active, the service is closed.

\subsubsection{Access Manager ($ACC$)}
The Access Manager $ACC$ is responsible for connecting to the
\gls{RAN} whenever a service $SRV\in\mathbf{SRV}(UE)$ requests it.
Its Atomic \gls{DEVS} model is as follows:
\begin{align*}
		ACC&=\langle X,S,Y,\delta _{int},\delta _{ext},\lambda,ta\rangle\\
		X&=\{in\_srv,in\_snr\}\\
		S&=\{\langle\sigma,\phi,\mathbf{snr},\mathbf{srvs},ap\rangle\}\\
		Y&=\{out\_srv,out\_connect,out\_connected,out\_disconnect\}\\
		ta&(\langle\sigma,\phi,\mathbf{snr},\mathbf{srvs},best\rangle)=\sigma\\
		\lambda &(\langle\sigma,\phi,\mathbf{snr},\mathbf{srvs},ap\rangle)=\begin{cases}
			(out\_srv,false),&\text{if }ap=\emptyset\\
			(out\_srv,true),&\text{otherwise}\\
		\end{cases}\\
		\lambda &(\langle\sigma,connect,\mathbf{snr},\mathbf{srvs},ap\rangle)=(out\_connect,ap))\\
		\lambda &(\langle\sigma,disconnect,\mathbf{snr},\mathbf{srvs},ap\rangle)=(out\_disconnect,ap)\\
		\lambda &(\langle\sigma,set,\mathbf{snr},\mathbf{srvs},ap\rangle)=(out\_connected,ap)\\
		\lambda &(\langle\sigma,change,\mathbf{snr},\mathbf{srvs},ap\rangle)=\{(out\_disconnect,ap),(out\_connect,best(\mathbf{snr})\}\\
		\delta _{int}&(\langle\sigma,\phi',\mathbf{snr},\mathbf{srvs},ap\rangle)=\langle\infty,on,\mathbf{cap},\mathbf{srvs},best(\mathbf{snr})\rangle\\
			&\hspace{2cm}\phi'\in\{connect,on,set,change\}\\
		\delta _{int}&(\langle\sigma,disconnect,\mathbf{snr},\mathbf{srvs},ap\rangle)=\langle\infty,off,\mathbf{snr},\mathbf{srvs},\emptyset\rangle\\
		\delta _{ext}&(\langle\sigma,\phi,\mathbf{snr},\mathbf{srvs},ap\rangle,e,(in\_st,\langle app, true\rangle))=\langle0,\phi,\mathbf{snr},\mathbf{srvs}\cup app,ap\rangle\\
		\delta _{ext}&(\langle\sigma,\phi,\mathbf{snr},\mathbf{srvs},ap\rangle,e,(in\_st,\langle app, false\rangle))=\langle0,\phi,\mathbf{snr},\mathbf{srvs}\setminus app,ap\rangle\\
		\delta _{ext}&(\langle\sigma,off,\mathbf{snr},\mathbf{srvs},\emptyset\rangle,e,(in\_snr,\langle AP,SNR\rangle))=\langle\infty,off,\mathbf{snr'},\mathbf{srvs},\emptyset\rangle\\
		\delta _{ext}&(\langle\sigma,on,\mathbf{snr},\mathbf{srvs},ap\rangle,e,(in\_snr,\langle AP,SNR\rangle))=\\
			&\hspace{4.5cm}=\begin{cases}
				\langle\infty,on,\mathbf{snr'},\mathbf{srvs},ap\rangle,&\text{if }best(\mathbf{snr'})\neq ap\\
				\langle0,change,\mathbf{snr'},\mathbf{srvs},ap\rangle,&\text{otherwise}
			\end{cases}\\
\end{align*}

Services $SRV\in\mathbf{SRV}(UE)$ notify whether they need a connection
or not via the $in\_srv$ port. $ACC$ keeps a record of ongoing services
in $\mathbf{srvs}$. If any service requires a connection,
$ACC$ proceeds to connect to the $AP\in\mathbf{AP}$ that offers the highest
\gls{SNR}, $ap$.
To do so, $ACC$ receives reports of the \gls{SNR}
perceived by the $UE$ for every $AP\in\mathbf{AP}$ via the $in\_snr$
port. $ACC$ keeps a record of $SNR(AP\in\mathbf{AP})$ in $\mathbf{snr}$.

So, when a service requires a connection, $ACC$ selects which $AP$ offers
the highest \gls{SNR}, and 
If a new \gls{SNR} is reported while $UE$ is connected to a given $AP$,
and the most suitable \gls{AP} $ap_{best}$ differs from the connected \gls{AP} $ap$, a handover process from $ap$ to $ap_{best}$ is triggered.

\subsubsection{Mobility Module ($MOB$)}
\gls{UE} devices are the only mobile components of the scenario. Every
$UE\in\mathbf{UE}$ contains a $MOB$ module that manages the location
of the \gls{UE}.

\subsubsection{Antenna ($ANT$)}
$ANT$ is the antenna used by the \gls{UE} for communicating
with the \glspl{AP} via wireless channels. Its behavior is similar to
radio antennas of \glspl{AP}. However, as \gls{UE} have more power
limitation than \glspl{AP}, transmitting power of \gls{UE}'s antennas
is limited to 30 \si{\decibelm}.

\subsection{Crosshaul Network ($XH$) and Radio Interface ($RAD$)}
The crosshaul network acts as the intercommunication interface of
all the elements connected to the \gls{RAN} (i.e., $\mathbf{AP}$,
$\mathbf{EDC}$, and $SDNC$). 
All the input/output interfaces of the $XH$ module accept physical
messages, which would be equivalent to light impulses through the
optical fiber links that interconnect the components.
The attributes that define a physical message $M$ are gathered in
Tab.~\ref{tab:msg}.
\begin{table}[h]
	\caption{Attributes of Physical Messages}
	\label{tab:msg}
	\centering
	\begin{tabu} to \linewidth {X[1,c,m] X[4.5,l,m]}
		\bfseries Attribute & \bfseries Definition\\
		\hline\hline
		$from(M)$ & Network node that sent the message.\\
		$to(M)$ & Network node that has to receive the message.\\
		$data(M)$ & Information to be sent from $from(M)$ to $to(M)$.\\
		$bw(M)$ & Bandwidth (in \si{\hertz}) used for sending the message\\
		$pw(M)$ & Power (in \si{\decibelm}) of the signal that carries the information to be sent.\\
		$eff(M)$ & Spectral efficiency (in
		\si{\bits\per\second\per\hertz}) used for sending the message.\\
		$s(M)$ & Size (in \si{\bits}) of the message.\\
		\hline
	\end{tabu}
\end{table}

$XH$ is defined as the following Coupled \gls{DEVS} model:
\begin{align*}
	&\hspace{1cm}\forall\thinspace EDC\in\mathbf{EDC},\forall\thinspace AP\in\mathbf{AP}:\\
		XH&=\langle X,Y,C,EIC,IC,EOC\rangle\\
		X&=\{in\_ul(EDC),in\_ul(AP),\\
			&\hspace{0.6cm}in\_dl(SDNC),in\_dl(EDC)\}\\
		Y&=\{out\_ul(SDNC),out\_ul(EDC),out\_dl(AP)\}\\
		C&=\{ULCH(EDC,SDNC),ULCH(AP,EDC),\\
			&\hspace{0.6cm}DLCH(EDC,AP),ULCH(SDNC,AP)\}\\
		EIC&=\{XH \xrightarrow[\text{in}]{\text{in\_ul(EDC)}} ULCH(EDC,SDNC), \\
			&\hspace{0.6cm} XH \xrightarrow[\text{in}]{\text{in\_ul(AP)}} ULCH(AP,EDC),\\
			&\hspace{0.6cm} XH \xrightarrow[\text{in}]{\text{in\_dl(SDNC)}} DLCH(SDNC,AP), \\
			&\hspace{0.6cm} XH \xrightarrow[\text{in}]{\text{in\_dl(EDC)}} DLCH(EDC,AP)\}\\
		IC&=\{\}\\
		EOC&=\{ULCH(EDC,SDNC) \xrightarrow[\text{out\_ul(SDNC)}]{\text{out}} XH, \\
			&\hspace{0.6cm} ULCH(AP,EDC) \xrightarrow[\text{out\_ul(EDC)}]{\text{out}} XH,\\
			&\hspace{0.6cm} DLCH(EDC,AP) \xrightarrow[\text{out\_dl(AP)}]{\text{out}} XH, \\
			&\hspace{0.6cm} DLCH(SDNC,AP) \xrightarrow[\text{out\_dl(AP)}]{\text{out}} XH\}
\end{align*}

The components $ULCH(X,Y)$ and $DLCH(Y,X)$ model two physical
channels (Uplink and Downlink, respectively) used for inter-module
communication. Together, they emulate the behavior of an \gls{FDD}
single-mode SMF-28 optical fiber communication links using the third
window (operating wavelength at 1550 \si{\nano\meter})~\citep{corning2014}.

On the other hand, the $RAD$ component models a radio interface. This
module is similar to $XH$, as its function is to provide
communication between \gls{UE} and \glspl{AP} – i.e., the radio
model behaves as the wireless communication channels between them,
and messages can be interpreted as electromagnetic waves propagated
through the radio. Its Coupled \gls{DEVS} model is:

\begin{align*}
	&\hspace{1cm}\forall\thinspace UE\in\mathbf{UE},\forall\thinspace AP\in\mathbf{AP}\\
		RAD&=\langle X,Y,C,EIC,IC,EOC\rangle\\
		X&=\{in\_pbch(AP),in\_pdsch(AP),in\_pdcch(AP),\\
			&\hspace{0.6cm}in\_pusch(UE),in\_pucch(UE)\}\\
		Y&=\{out\_pbch(UE),out\_pdcch(UE),\\
			&\hspace{0.6cm}out\_pdsch(UE),out\_pucch(AP),\\
			&\hspace{0.6cm}out\_pusch(AP)\}\\
		C&=\{PBCH(AP,UE),PUCCH(UE,AP),\\
				&\hspace{0.6cm}PDCCH(AP,UE),PUSCH(UE,AP),\\
				&\hspace{0.6cm}PDSCH(AP,UE)\}\\
		EIC&=\{RAD \xrightarrow[\text{in}]{\text{in\_pbch(AP)}} PBCH(AP,UE), \\
			&\hspace{0.6cm}RAD \xrightarrow[\text{in}]{\text{in\_pdcch(AP)}} PDCCH(AP,UE),\\
			&\hspace{0.6cm}RAD \xrightarrow[\text{in}]{\text{in\_pdsch(AP)}} PDSCH(AP,UE),\\
			&\hspace{0.6cm}RAD \xrightarrow[in]{in\_pucch(UE)} PUCCH(UE,AP),\\
			&\hspace{0.6cm}RAD \xrightarrow[in]{in\_pusch(UE)} PUSCH(UE,AP)\}\\
		IC&=\{\}\\
		EOC&=\{PBCH(AP,UE) \xrightarrow[out\_pbch(UE)]{out} RAD ,\\
			&\hspace{0.6cm}PDCCH(AP,UE)\xrightarrow[out\_pdcch(UE)]{out} RAD ,\\
			&\hspace{0.6cm}PDSCH(AP,UE)\xrightarrow[out\_pdsch(UE)]{out} RAD ,\\
			&\hspace{0.6cm}PUCCH(UE,AP)\xrightarrow[out\_pucch(AP)]{out} RAD ,\\
			&\hspace{0.6cm}PUSCH(UE,AP)\xrightarrow[out\_pusch(AP)]{out} RAD \}
\end{align*}
$PBCH$ corresponds to the \gls{PBCH}, which is used by \glspl{AP} for
broadcasting information to all the \gls{UE} nodes of the scenario.
$PUCCH$ and $PDCCH$ correspond to the \gls{PUCCH} and \gls{PDCCH},
respectively. They constitute an \gls{FDD} logical communication
channel between \gls{UE} and the \gls{AP} they are connected to.
These channels are used to send control messages related to
\gls{UE}' connectivity.
$PUSCH$ and $PDSCH$ model the \gls{PUSCH} and \gls{PDSCH}, which
constitute another \gls{FDD} logical channel between \glspl{AP} and
\gls{UE}. These physical channels are used exclusively for sending
data related to the \gls{IoT} applications running on the \gls{UE} devices.

The atomic \gls{DEVS} models of communication channels is:
\begin{align*}
		CH&(FROM,TO)=\langle X,S,Y,\delta _{int},\delta _{ext},\lambda,ta\rangle\\
		X&=\{in\}\\
		S&=\{\langle\sigma,next,\mathbf{q}\rangle\}\\
		Y&=\{out\}\\
		ta&(\langle\sigma,next,\mathbf{q}\rangle)=\sigma\\
		\lambda &(\langle\sigma,next,\mathbf{q}\rangle)=\{(out,next)\}\\
		\delta _{int}&(\langle\sigma,next,\mathbf{q}\neq\emptyset\rangle)=\langle del(M),att(M),\mathbf{q}\setminus M\rangle\\
		&\hspace{3.4cm}M=\mathbf{q}\{0\}\\
		\delta _{int}&(\langle\sigma,next,\emptyset\rangle)=\langle\infty,\emptyset,\emptyset\rangle\\
		\delta _{ext}&(\langle\sigma,next\neq\emptyset,\mathbf{q}\rangle,e,(in,M))=\langle\sigma - e,next,\mathbf{q}\leftarrow M\rangle\\
		\delta _{ext}&(\langle\infty,\emptyset,\mathbf{q}\rangle,e,(in,M))=\langle 0,\emptyset,\mathbf{q}\leftarrow M\rangle\\
\end{align*}

When a new physical message $M$ is received by the channel via the
$in$ port, $M$ is appended at the end of a sorted sequence of messages,
$\mathbf{q}$. If no message is being sent, the first message of the
queue is marked as the $next$ to be sent. To emulate the effect of
attenuation, $pw(next)$ is reduced by the power budget, $L$.
The attenuation of $XH$ channels is considered negligible, as distances
between nodes within a \gls{RAN} is small enough to be considered with
active optical fiber links.
In case of the channels that compose the radio interface, the
attenuation function is modeled as the \gls{FSPL} function:
\begin{equation}\label{eq:delay}
	L=20\log_{10}{\left(\frac{4\pi d f}{c}\right)}\thinspace\lbrack\si{\decibel}\rbrack
\end{equation}
where $d$ is the distance in \si{\meter} between the nodes, and
$f$ is the carrier frequency in \si{\hertz} used for transmitting
the message. We set $f$ to 33 \si{\giga\hertz}, which is the lowest
available carrier frequency for the radio band n77~\citep{huaweispectrum}.

For emulating the delay introduced by the network, the channel sets
$\sigma$ is set to $del(M)$:
\begin{equation}
	del(M)=D_{tr}+D_{prop}=\frac{s(M)}{eff(M) \times bw(M)}+\frac{d}{V_{prop}}
\end{equation}
$D_{tr}$ corresponds to the transmission delay. It depends on the
size of the message $s(M)$ (in \si{\bits}), the bandwidth used for
transmitting it $bw(M)$ (in \si{\hertz}), and the spectral efficiency
applied $eff(M)$ (in \si{\bits\per\second\per\hertz}).
All the messages sent through the crosshaul use a bandwidth of 10
\si{\giga\hertz} and a spectral efficiency of 1
\si{\bits\per\second\per\hertz}. Messages sent through the radio
interface vary both parameters, depending on the decisions made by \glspl{AP}
Alternatively, $D_{prop}$ corresponds to the propagation delay. It
depends on the distance between the nodes $d$ in \si{\meter} and the
propagation speed of the physical means $V_{prop}$. Again, $D_{prop}$ is negligible compared to $D_{tr}$.



%% file: text/5_simulation.tex
The scenario under study consists of an incremental learning
\gls{ADAS} use case. Each vehicle executes an \gls{ML} model
for detecting potential hazardous situations. 
Vehicles gather new data at a rate of 1
\si{\mega\bits\per\second}. Datastream is sent to the
corresponding \gls{EDC} in charge of the computation
offloading, which consists of an incremental learning
service for each vehicle.
To do so, \glspl{EDC} inject in real-time the new images to
the training process of the predictive models. Once they
significantly outperform the onboard versions, the embarked
models can be upgraded on the fly.

Each \gls{EDC} comprises 10 \glspl{PU} based on the AMD Sapphire
Pulse Radeon RX 580 series \gls{GPU}.
Resource and power consumption models for the \glspl{PU} correspond to
state-of-the-art models~\citep{perez2019}. Training a given \gls{UE}'s
predictive model requires 20\% of the computing resources of one
\gls{GPU}. \glspl{GPU} power consumption model is based on an
\gls{ANN} that depends on the \gls{ADAS} workload.
On the other hand, timing properties for the \glspl{PU} were chosen to
illustrate their impact on the outcome of the mathematical
model presented in this paper.
In this way, adding new images to the training data set does not
introduce any additional delay, as it is performed in the background.
However, \glspl{GPU} will add a delay of 0.2 seconds when opening
or closing a session. Furthermore, when an \gls{EDC} resource manager
commands a \gls{GPU} to switch on/off, the \gls{GPU} will be busy for
1 second before being able to perform any other activity (as to activate a \gls{GPU} from a deep sleep state). Note that
these parameters are merely an example and can be reconfigured as
desired, enabling modelers to explore different scenarios effortlessly.

\begin{figure}[h]
	\centering
	\includegraphics[width=0.7\linewidth]{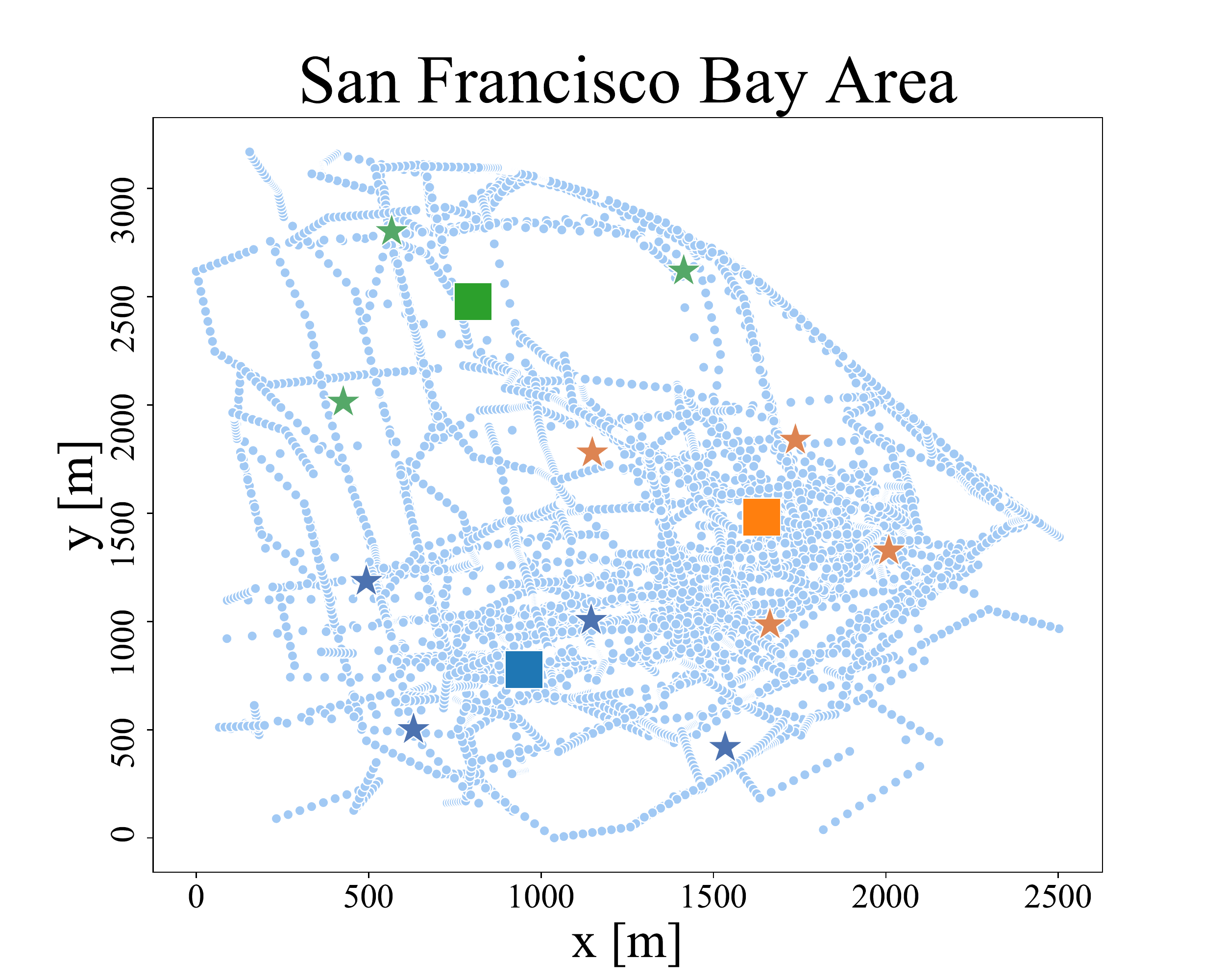}
	\caption{Use Case Scenario.}
	\label{fig:frisco}
\end{figure}

For simulating the scenarios, we used Mercury~\citep{Cardenas2019},
a \gls{MSO} framework for edge computing, which model relies
on \gls{DEVS} and complies the mathematical approach described in this
research.
The scenario, shown in \figurename~\ref{fig:frisco}, is based on real
mobility traces of taxis in the San Francisco bay
area~\citep{comsnets09piorkowski}.
Each trace corresponds to a GPS location of a San Francisco
based Yellow Cab vehicle. The dataset contains four features: latitude,
longitude, occupancy, and time. Latitude and longitude are in decimal degrees,
occupancy shows if a cab has a fare (1 = occupied, 0 = free), and time is in UNIX
epoch format.

A total of 220 simulations were performed. We explored different resource
management strategies on the \glspl{EDC} and the impact of these
strategies on the delay perceived by \gls{UE} and the power consumption
required by \glspl{EDC} depending on the number of \gls{UE} devices in the
scenario. Two dispatching algorithms were defined for the Resource Manager:
\begin{subnumcases}{PU_{next}(SRV)=}
	\underset{PU\in\mathbf{PU}(EDC)}{\text{arg min}} & $\frac{u(PU)}{U(PU)}$ \label{eq:emptiest}\\
	\hspace{0.5cm}st.& u(PU)+U(SRV) $\leq$ U(PU)\nonumber\\
	\underset{PU\in\mathbf{PU}(EDC)}{\text{arg max}} & $\frac{u(PU)}{U(PU)}$ \label{eq:fullest}\\
	\hspace{0.5cm}st.&u(PU)+U(SRV) $\leq$ U(PU)\nonumber
\end{subnumcases}
Dispatching algorithm defined in Eq.~(\ref{eq:emptiest}) assigns new
incremental learning services to the emptiest \gls{PU}.
On the other hand, the algorithm of Eq.~(\ref{eq:fullest}) selects the
\gls{PU} with less available computing resources providing that the
\gls{PU} has enough available resources.
The number of \glspl{PU} in hot standby for each \gls{EDC}, $N_{STBY}$,
ranged from 0 (i.e., all the \glspl{PU} are powered
off if no service is ongoing) to 10 (i.e., all the \glspl{PU} are
always powered on).

We executed the simulations on a MacBook Pro Retina, 15-inch Mid 2015, 2.5
Quad-Core Intel Core i7 with 16 GB 1600 MHz DDR3 memory, using the PyCharm
2020.1.1 IDE in sequential mode.
The simulation time increased following a complexity order of $O(n^2)$ as the
number of \gls{UE} devices grew. For example, scenarios with 10 \gls{UE} devices took a mean
time of 33 seconds, whereas more crowded scenarios with 100 \gls{UE} nodes took a
mean time of 22 minutes.
The resource allocation policy implemented by the
\glspl{EDC} did not impact on the simulation performance, as both algorithms have
a complexity order of $O(n)$, where $n$ is the number of \glspl{PU} in the
\gls{EDC}. The only difference between both policies is that one attempts to
minimize the cost function, whereas the other maximizes it.

\begin{figure}[h]
	\centering
	\includegraphics[width=0.9\textwidth]{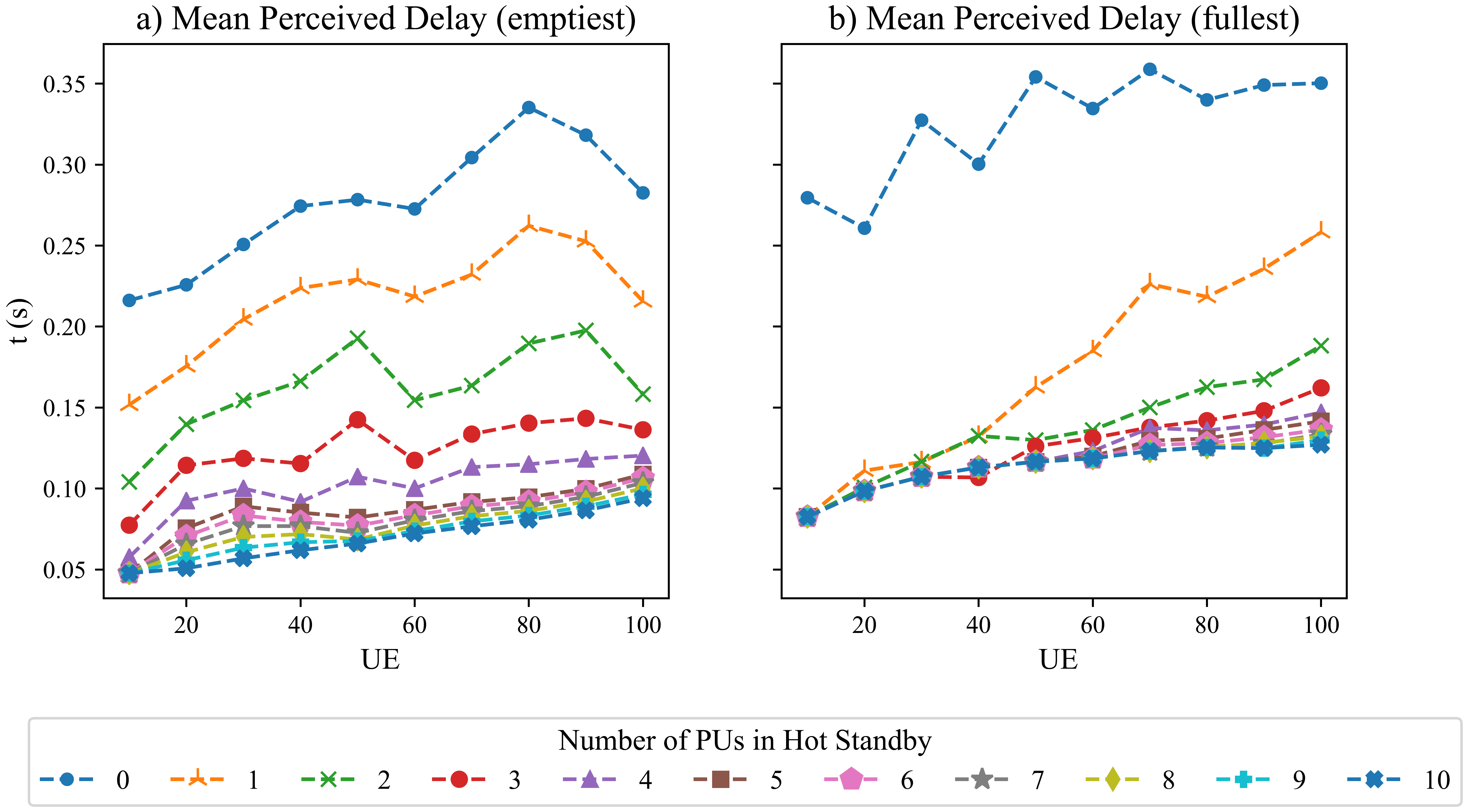}
	\caption{Delay Simulation Results.}
	\label{fig:res_delay}
\end{figure}

\figurename~\ref{fig:res_delay} depicts the obtained results regarding the
delay perceived by end users. \figurename~\ref{fig:res_delay}a corresponds to
those scenarios in which \glspl{EDC} implement the dispatching policy described
in Eq.~(\ref{eq:emptiest}), which assigns new sessions to the emptiest \gls{PU}
(i.e., the \gls{PU} with more available resources). On the other hand,
\glspl{EDC} of the scenarios shown in \figurename~\ref{fig:res_delay}b used the
dispatching policy described in Eq.~(\ref{eq:fullest}), which assigns new
sessions to the fullest \gls{PU} (i.e., the \gls{PU} with less available
resources).

In general, the delay is proportional to the \gls{UE} density. As the number of
connected devices increments, the available bandwidth assigned to each
\gls{UE} is less, and transmission delay of data packages is higher as well
(see Eq.~(\ref{eq:delay})).
On the other hand, increasing the number of \glspl{PU} in hot standby reduces
the delay perceived by \gls{UE}, because the probability of selecting a
switched off processing unit to host a new service decreases.
Regarding the dispatching algorithm, the one described in
Eq.~(\ref{eq:emptiest}), shown in \figurename~\ref{fig:res_delay}a,
tends to experience less delay than the algorithm of Eq.~(\ref{eq:fullest})
(in \figurename~\ref{fig:res_delay}b).
This difference is due to a higher probability when applying the second algorithm
of a new service being assigned to a busy \gls{PU}, thus driving to aggregated
delays for starting a service.

However, the second algorithm shows better delay reduction when combining it with
hot standby policies. In fact, for scenarios with low \gls{UE} density,
using only one or two \glspl{PU} in hot standby shows a mean delay similar to
keeping all the \glspl{PU} always switched on. On the other hand, the dispatching
policy that selects the emptiest \gls{PU} to host new services needs four or more
\glspl{PU} to obtain a satisfactory delay improvement, and this improvement never
equals keeping all the \glspl{PU} switched on.

\begin{figure}[h]
	\centering
	\includegraphics[width=0.9\textwidth]{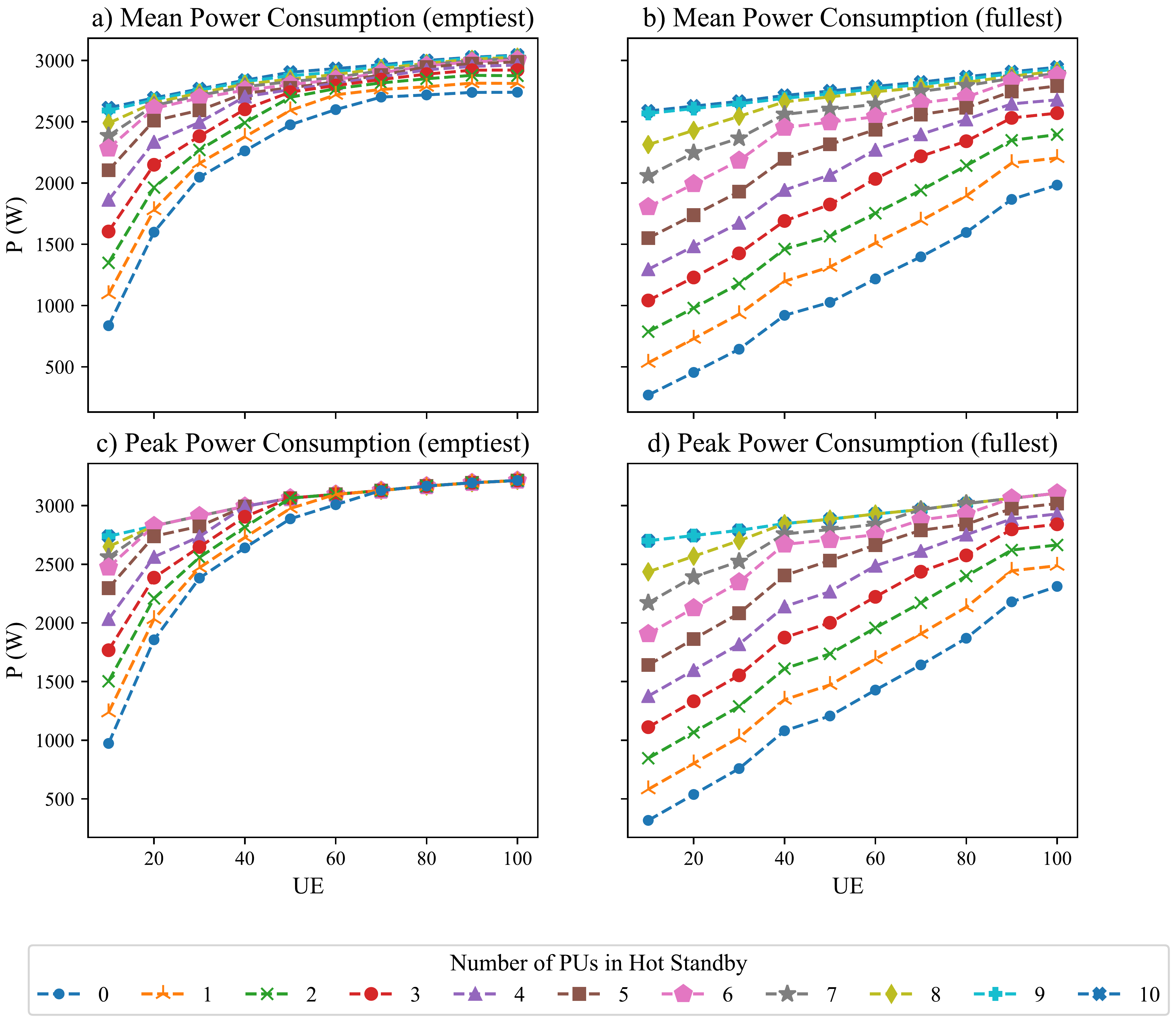}
	\caption{Mean and Peak Power Consumption Simulation Results.}
	\label{fig:res_power}
\end{figure}

Simulation results about the power consumption of the edge computing
infrastructure are shown in \figurename~\ref{fig:res_power}. Figures
\ref{fig:res_power}a and \ref{fig:res_power}b represent the mean power
consumption of the edge federation, whereas Figures \ref{fig:res_power}c and
\ref{fig:res_power}d show the peak power consumption.
The plots of the first column (i.e., Figures \ref{fig:res_power}a and
\ref{fig:res_power}c) contain information about the scenarios in which
\glspl{EDC} implement the dispatching policy described in
Eq.~(\ref{eq:emptiest}). On the other hand, \glspl{EDC} of the scenarios shown in
Figures \ref{fig:res_power}b and \ref{fig:res_power}d used the dispatching policy
described in Eq.~(\ref{eq:fullest}).

Mean and peak power consumption of the edge infrastructure are proportional to
the \gls{UE} density. This is due to the higher usage of computation offloading
resources.
If we focus on hot standby policies, increasing the number of \glspl{PU} in hot
standby affects negatively to the overall power consumption, as there are more
\glspl{PU} idling (i.e., switched on even if no services are active).
Comparing the effect of the dispatching algorithms, the policy that assigns new
sessions to the fullest \gls{PU} reported significantly less power consumption
(see Figures \ref{fig:res_power}b and \ref{fig:res_power}d).

Again, the second dispatching algorithm showed better results when combining it
with hot standby policies. If we focus on the results reported by the first
algorithm, the mean power consumption quickly converges to 2,500 Watts. Looking
at \figurename~\ref{fig:res_power}c, the peak power consumption reported by the
first dispatching algorithm for scenarios with 60 or more \gls{UE} is almost the
same, regardless of the number of \glspl{PU} in hot standby. On the other hand,
using the second dispatching algorithm, power consumption can be significantly
reduced using a low number of \glspl{PU} in hot standby, and this power saving is
achieved even in scenarios with a high \gls{UE} density.

The following conclusions can be drawn from the obtained results:
The dispatching algorithm described in Eq.~(\ref{eq:fullest})
offers a better compromise between delay and power
consumption; the delay is kept low when using a reduced number of
\glspl{PU} in hot standby. However, if the \gls{IoT} application
is very sensitive to delay, the algorithm described in
Eq.~(\ref{eq:emptiest}) is able to outperform the other
algorithm at expenses of increasing the number of required
\glspl{PU} in hot standby and, hence, the overall power consumption.

%% file: text/6_conclusions.tex
This research presents a formal \gls{MS} specification focused
on deploying effective \gls{IoT} architectures supported by
computation offloading on the edge of the network.
Edge computing infrastructures will be key technology enablers
for boosting the performance of new \gls{IoT} services that rely
on \gls{AI} and \gls{ML} algorithms, providing a sustainable
and scalable  approach for applying incremental learning
methodologies and improving in real-time service performance.

System specification was carried out following an \gls{MBSE}
approach together with the \gls{DEVS} mathematical formalism.
The model describes a complete architecture of edge scenarios,
from \gls{UE} to \glspl{PU}, and provides information about
intercommunication between entities, latency perceived by
\gls{UE} devices, and operational expenditure of edge infrastructures.
Location awareness, an inherent characteristic of edge
computing, is gathered in the model, providing real-time
dynamic infrastructure re-configuration.

Using complete, formal \gls{MS} tools like the one described in
this research assists in the decision-making process required
for the deployment of edge computing infrastructure.
We exemplify its benefits by presenting a use case based on
the incremental learning of \gls{ADAS} applications. We explore
different resource management policies using simulation
tools compliant with the model under study.
Simulation results are discussed, and the most suitable
configurations are identified, taking into account both
\gls{QoS} and power consumption.

\subsection{Future Work}
Regarding future directions, we are currently introducing
temperature models for \glspl{PU}, and \glspl{EDC} cooling
power models. Cooling supposes a significant power
consumption share in state-of-the-art data centers.
Smart grid modules will be defined to simulate
renewable energy source and storage systems. Adding these
new contributions will enable us to explore other
causes of operational expenses.

Finally, we will add cloud storage to the incremental learning
process. Standard \gls{ML} models are stored in the cloud.
As new predictive models are obtained from the computation 
offloading, these models will not only be transferred to
\gls{UE} but also to the cloud for increasing availability
and data persistency.

%% file: text/ack.tex
This project has been partially supported by the Centre for the Development of
Industrial Technology (CDTI) under contracts IDI-20171194, IDI-20171183 and RTC-2017-6090-3
and by the Education and Research Council of the Community of Madrid (Spain), under research grant S2018/TCS-4423.

%% file: text/annex_devs.tex
\gls{DEVS} is a general formalism for discrete event system modeling based on set theory~\citep{zeigler2000}. The \gls{DEVS} formalism provides the framework for information modeling which gives several advantages to analyze and design complex systems: completeness, verifiability, extensibility, and maintainability. Once a system is described in terms of the \gls{DEVS} theory, it can be easily implemented using an existing computational library. The parallel \gls{DEVS} (PDEVS) approach was introduced, after 15 years, as a revision of Classic \gls{DEVS}. Currently, PDEVS is the prevalent DEVS, implemented in many libraries. In our work, unless it is explicitly noted, the use of \gls{DEVS} implies PDEVS.

\gls{DEVS} enables the representation of a system by three sets and five functions: input set $(X)$, output set $(Y)$, state set $(S)$, external transition function ($\delta_{\rm ext}$), internal transition function ($\delta_{\rm int}$), confluent function ($\delta_{\rm con}$), output function ($\lambda$), and time advance function ($ta$). 

DEVS models are of two types: atomic and coupled. Atomic models are directly expressed in the DEVS formalism specified above. Atomic DEVS processes input events based on their model's current state and condition, generates output events and transition to the next state. The coupled model is the aggregation/composition of two or more atomic and coupled models connected by explicit couplings. Particularly, an atomic model is defined by the following equation:

\begin{equation}
A=\langle X, Y, S, \delta_{\rm ext},  \delta_{\rm int}, \delta_{\rm con}, \lambda, ta\rangle
\label{eq:DevsAtomicModel}
\end{equation}

\noindent where:

\begin{itemize}

\item $X$ is the set of inputs described in terms of pairs port-value: $\left\{ p \in IPorts,v \in X_p \right\} $.

\item $Y$ is the set of outputs, also described in terms of pairs port-value: $\left\{ p \in OPorts,v \in Y_p \right\} $.

\item $S$ is the set of sequential states.

\item $\delta_\mathrm{ext}: Q\times X^b \rightarrow S$ is the external transition function. It is automatically executed when an external event arrives to one of the input ports, changing the current state if needed.
\begin{itemize}
\item $Q={(s,e) s \in S, 0 \leq e \leq ta(s)}$ is the total state set, where $e$ is the time elapsed since the last transition.
\item $X^b$ is the set of bags over elements in $X$.
\end{itemize}

\item $\delta_\mathrm{int}: S \rightarrow S$ is the internal transition function. It is executed right after the output ($\lambda$) function and is used to change the state $S$.

\item $\delta_\mathrm{con}: Q\times X^b \rightarrow S$ is the confluent function, subject to $\delta_\mathrm{con}(s,ta(s),\emptyset)=\delta_\mathrm{int}(s)$. This transition decides the next state in cases of collision between external and internal events, i.e., an external event is received and elapsed time equals time-advance. Typically, $\delta_\mathrm{con}(s,ta(s),x) = \delta_\mathrm{ext}(\delta_\mathrm{int}(s),0,x)$.

\item $\lambda: S \rightarrow Y^b$ is the output function. $Y^b$ is the set of bags over elements in $Y$. When the time elapsed since the last output function is equal to $ta(s)$, then $\lambda$ is automatically executed.

\item $ta(s): S \rightarrow \Re_0^+ \cup \infty$ is the time advance function.
\end{itemize}

An atomic model's state is $s\in S$ at any given time $t$. If no external events
occur, its state remains in $s$ for a period of time $ta(s)$ (i.e., time advance
function). When the lifetime expires, the atomic model sends a set of output
events $Y^b\in Y$ according to its output function $\lambda(s)$, and changes its
state to a new one given by the internal transition function $\delta_{int}(s)$.
If one or more input events $X^b\in X$ occur before the expiration of $ta(s)$,
the model changes to a new state determined by the external transition function
$\delta_{ext}(s,e,X^b)$. The confluent transition function $\delta_{con}$
determines the next state in the case of collisions when a model receives
external events at the same time of its internal transition.

The formal definition of a coupled model is described as:
\begin{equation}
M = \langle X, Y, C, EIC, EOC, IC \rangle
\end{equation}

\noindent where:
\begin{itemize}
\item $X$ is the set of inputs described in terms of pairs port-value: $\left\{ p \in IPorts,v \in X_p \right\} $.
\item $Y$ is the set of outputs, also described in terms of pairs port-value: $\left\{ p \in OPorts,v \in Y_p \right\} $.
\item $C$ is a set of DEVS component models (atomic or coupled). Note that $C$ makes this definition recursive.
\item $EIC$ is the external input coupling relation, from external inputs of $M$ to component inputs of $C$.
\item $EOC$ is the external output coupling relation, from component outputs of $C$ to external outputs of $M$.
\item $IC$ is the internal coupling relation, from component outputs of $c_i \in C$ to component outputs of $c_j \in C$, provided that $i \neq j$.
\end{itemize}
Given the recursive definition of $M$, a coupled model can itself be a part of a component in a larger coupled model system giving rise to a hierarchical \gls{DEVS} model construction.